\documentclass[zpreprint,zbstepj]{./LaTeX/zeus/zeus_paper}
\usepackage[english]{babel}

\newcommand{\ZcoosysB}{%
The ZEUS coordinate system is a right-handed Cartesian system, with the $Z$
axis pointing in the proton beam direction, referred to as the ``forward
direction'', and the $X$ axis pointing left towards the centre of HERA.
The polar angle, $\theta$, is measured with respect to the proton beam
direction. The coordinate origin is at the nominal interaction point.\xspace}
\newcommand{\Zpsrap}{%
The pseudorapidity is defined as $\eta=-\ln\left(\tan\frac{\theta}{2}\right)$,
where the polar angle, $\theta$, is measured with respect to the proton beam
direction.\xspace}

\newcommand{\ZcoosysfnB}{\footnote{\ZcoosysB}}

\newcommand{\ZcoosysfnBeta}{\footnote{\ZcoosysB\Zpsrap}}

\newcommand{\Zctddesc}[1]{%
Charged particles are tracked in the central tracking detector (CTD)~\citeCTD,
which operates in a magnetic field of $1.43\Tesla$ provided by a thin 
superconducting solenoid. The CTD consists of 72~cylindrical drift chamber 
layers, organised in nine~superlayers covering the polar-angle\ZcoosysfnBeta~region 
\mbox{$15^\circ<\theta<164^\circ$}. The relative transverse-momentum 
resolution for
full-length tracks is $\sigma(p_T)/p_T=0.0058p_T\oplus0.0065\oplus0.0014/p_T$,
with $p_T$ in $\Gev$.\xspace\\
In 2001 a silicon microvertex detector (MVD)~\cite{nim:a453:89,nim:a505:663} was installed 
inside the CTD. The MVD is organised into a barrel with 3 cylindrical layers and a forward 
section with four planar layers perpendicular to the HERA beam direction. The barrel 
contains 600 single-sided silicon strip sensors each having 512 120 $\rm{\mu m}$ wide strips; the 
forward section contains 112 sensors each of which has 480 120 $\rm{\mu m}$ wide strips.\xspace\\}

\chardef\usc=95
\chardef\til=126
\catcode`\@=11 
\DeclareRobustCommand\xdotspace{\futurelet\@let@token\@xdotspace}
\def\@xdotspace{%
  \ifx\@let@token.\else
  \ifx\@let@token\bgroup.\else
  \ifx\@let@token\egroup.\else
  \ifx\@let@token\/.\else
  \ifx\@let@token\ .\else
  \ifx\@let@token~.\else
  \ifx\@let@token!.\else
  \ifx\@let@token,.\else
  \ifx\@let@token:.\else
  \ifx\@let@token;.\else
  \ifx\@let@token?.\else
  \ifx\@let@token/.\else
  \ifx\@let@token'.\else
  \ifx\@let@token).\else
  \ifx\@let@token-.\else
  \ifx\@let@token\@xobeysp.\else
  \ifx\@let@token\space.\else
  \ifx\@let@token\@sptoken.\else
   .\space
   \fi\fi\fi\fi\fi\fi\fi\fi\fi\fi\fi\fi\fi\fi\fi\fi\fi\fi}
\catcode`\@=12 

\newcommand{\stru}[2]{%
   \relax\ifmmode\hbox{\vrule height#1 depth#2 width0pt}%
   \else\vrule height#1 depth#2 width0pt\fi}

\newcommand{\Ronum}[1]{\uppercase\expandafter{\romannumeral#1}}
\newcommand{\ronum}[1]{\expandafter{\romannumeral#1}}
\DeclareRobustCommand{\LaTeXZ}{%
  \LaTeX\kern-.05em4\kern-.1em
  {\raisebox{-0.2ex}{$\scriptstyle\text{ZEUS}$}}\xspace}

\DeclareMathAlphabet{\mathbf}{OT1}{cmr}{bx}{sl}
\newcommand{\eVdist}{\kern-0.06667em}

\newcommand{\Gev}{{\text{Ge}\eVdist\text{V\/}}}

\newcommand{\gev}{{\,\text{Ge}\eVdist\text{V\/}}}

\newcommand{\pb}{\,\text{pb}}

\newcommand{\Tesla}{\,\text{T}}

\newcommand{\slashfrac}[2]{%
  \raisebox{0.5ex}{\ensuremath #1}\kern-0.12em/\kern-0.08em
  \raisebox{-.8ex}{\ensuremath #2}}

\newcommand{\sqr}[3]{%
    {\vcenter{\hrule height.#3ex\hbox{\vrule width.#2ex height#1ex
     \kern#1ex\vrule width.#3ex}\hrule height.#2ex}}}

\catcode`\@=11 
\newcommand{\parenbar}{\mathpalette\p@renb@r}
\def\p@renb@r#1#2{\vbox{%
  \ifx#1\scriptscriptstyle \dimen@.7em\dimen@ii.2em\else
  \ifx#1\scriptstyle \dimen@.8em\dimen@ii.25em\else
  \dimen@1em\dimen@ii.4em\fi\fi \offinterlineskip
  \ialign{\hfill##\hfill\cr
    \vbox{\hrule width\dimen@ii}\cr
    \noalign{\vskip-.3ex}%
    \hbox to\dimen@{$\mathchar300\hfil\mathchar301$}\cr
    \noalign{\vskip-.3ex}%
    $#1#2$\cr}}}
\catcode`\@=12 

\newcommand{\MSbar}{\hbox{$\overline{\rm MS}$}\xspace}

\newcommand{\IP}{{\rm I$\kern-0.01667em$P}\xspace}

\mathchardef\qsm=63
\mathchardef\pls=43
\mathchardef\mns=512
\mathchardef\plm=518
\mathchardef\eql=61
\mathchardef\smallleft=300
\mathchardef\smallright=301
\mathchardef\les=316
\mathchardef\gre=318
\mathchardef\leq=532
\mathchardef\grq=533

\catcode`\@=11 
\newcounter{pict@width}
\newcounter{pict@height}
\newlength{\pict@scale}
\newcommand{\psfigadd}[4]{%
\setcounter{pict@width}{1*\ratio{#2+\pict@scale/2}{\pict@scale}}
\setcounter{pict@height}{1*\ratio{#3+\pict@scale/2}{\pict@scale}}
\setlength{\unitlength}{\pict@scale}
\hbox to #2{\hspace{-\fill}\begin{picture}(\thepict@width,\thepict@height)
\put(0,0){\psfig{figure=#1,width=#2,height=#3,clip=}}
\SetScale{0.283466457}
\SetWidth{1.763889}
{#4}
\end{picture}}
}
\newcounter{pict@widthfst}
\newcounter{pict@widthscd}
\newcounter{pict@widthtot}
\newcommand{\psfigaddtwo}[7]{%
\setcounter{pict@widthfst}{1*\ratio{#2+\pict@scale/2}{\pict@scale}}
\setcounter{pict@widthscd}{1*\ratio{#2+#4+\pict@scale/2}{\pict@scale}}
\setcounter{pict@widthtot}{1*\ratio{#2+#4+#6+\pict@scale/2}{\pict@scale}}
\setcounter{pict@height}{1*\ratio{#3+\pict@scale/2}{\pict@scale}}
\setlength{\unitlength}{\pict@scale}
\hbox{\hspace{-\fill}\begin{picture}(\thepict@widthtot,\thepict@height)
\put(0,0){\psfig{figure=#1,width=#2,height=#3,clip=}}
\put(\thepict@widthscd,0){\psfig{figure=#5,width=#6,height=#3,clip=}}
\SetScale{0.283466457}
\SetWidth{1.763889}
{#7}
\end{picture}}
}
\newcommand{\psfigror}[4]{%
\setcounter{pict@width}{1*\ratio{#2+\pict@scale/2}{\pict@scale}}
\setcounter{pict@height}{1*\ratio{#3+\pict@scale/2}{\pict@scale}}
\setlength{\unitlength}{\pict@scale}
\hbox{\begin{picture}(\thepict@width,\thepict@height)
\put(0,\thepict@height){\psfig{figure=#1,width=#3,height=#2,clip=,angle=270}}
\SetScale{0.283466457}
\SetWidth{1.763889}
{#4}
\end{picture}}
}
\newcommand{\psfigrol}[4]{%
\setcounter{pict@width}{1*\ratio{#2+\pict@scale/2}{\pict@scale}}
\setcounter{pict@height}{1*\ratio{#3+\pict@scale/2}{\pict@scale}}
\setlength{\unitlength}{\pict@scale}
\hbox{\begin{picture}(\thepict@width,\thepict@height)
\put(0,0){\psfig{figure=#1,width=#3,height=#2,clip=,angle=90}}
\SetScale{0.283466457}
\SetWidth{1.763889}
{#4}
\end{picture}}
}
\catcode`\@=12 
\newlength\listtextwidth

\catcode`\@=11 
\newlength{\@tabfninsert}
\newlength{\@tabfnwidth}
\newcommand{\tabfootnote}[2]{%
  \setlength{\@tabfninsert}{0.8em}
  \setlength{\@tabfnwidth}{\textwidth}
  \addtolength{\@tabfnwidth}{-\@tabfninsert}
  \addtolength{\@tabfnwidth}{-0.4em}
  \noindent\makebox[\@tabfninsert][r]{\footnotesize$^{#1}$\hfil}\hfill%
  \parbox[t]{\@tabfnwidth}{\footnotesize #2\hfill}}
\catcode`\@=12 

\newcommand{\PTM}       {P_{T,{\rm miss}}}

\def\citeCTD{{\cite{%
nim:a279:290,*npps:b32:181,*nim:a338:254%
}}\xspace}
\def\citeCAL{{\cite{%
nim:a309:77,*nim:a309:101,*nim:a321:356,*nim:a336:23%
}}\xspace}

\begin{document}
\title{
Measurement of charged current deep inelastic scattering 
cross sections with a longitudinally polarised electron beam at HERA
}                                                       
                    
\author{ZEUS Collaboration}

\prepnum{DESY-08-177}

\abstract{
Measurements of the cross sections for charged current 
deep inelastic scattering in $e^{-}p$ collisions with 
longitudinally polarised electron beams are presented. 
The measurements are based on a data sample with an integrated luminosity 
of $175\pb^{-1}$ collected with the ZEUS detector at HERA 
at a centre-of-mass energy of 318\gev. 
The total cross section is given for positively and negatively 
polarised electron beams.
The differential cross-sections $d\sigma/d Q^2$, $d\sigma/d x$ and 
$d\sigma/d y$  are presented for \mbox{$Q^{2}>200\gev^2$}.
The double-differential cross-section $d^2 \sigma/d x d Q^2$ is presented 
in the kinematic range \mbox{$280<Q^2<30\,000 \gev^2$} and
 \mbox{$0.015<x<0.65$}.
The measured cross sections are compared with the predictions of 
the Standard Model.
}

\makezeustitle

\def\3{\ss}                                                                                        
\pagenumbering{Roman}                                                                              
                                                   %
\begin{center}                                                                                     
{                      \Large  The ZEUS Collaboration              }                               
\end{center}                                                                                       
  S.~Chekanov,                                                                                     
  M.~Derrick,                                                                                      
  S.~Magill,                                                                                       
  B.~Musgrave,                                                                                     
  D.~Nicholass$^{   1}$,                                                                           
  \mbox{J.~Repond},                                                                                
  R.~Yoshida\\                                                                                     
 {\it Argonne National Laboratory, Argonne, Illinois 60439-4815, USA}~$^{n}$                       
\par \filbreak                                                                                     
  M.C.K.~Mattingly \\                                                                              
 {\it Andrews University, Berrien Springs, Michigan 49104-0380, USA}                               
\par \filbreak                                                                                     
  P.~Antonioli,                                                                                    
  G.~Bari,                                                                                         
  L.~Bellagamba,                                                                                   
  D.~Boscherini,                                                                                   
  A.~Bruni,                                                                                        
  G.~Bruni,                                                                                        
  F.~Cindolo,                                                                                      
  M.~Corradi,                                                                                      
\mbox{G.~Iacobucci},                                                                               
  A.~Margotti,                                                                                     
  R.~Nania,                                                                                        
  A.~Polini\\                                                                                      
  {\it INFN Bologna, Bologna, Italy}~$^{e}$
% ...
\newpage                                                                                           
                                                           %
$^{\    1}$ also affiliated with University College London,                                        
United Kingdom\\                                                                                   
$^{\    2}$ now at University of Salerno, Italy \\                                                 
% ...
$^{  30}$ now at KEK, Tsukuba, Japan \\                                                            
% ...
%
\newpage   
                                                           %
                                                           %
\begin{tabular}[h]{rp{14cm}}                                                                       
$^{a}$ &  supported by the Natural Sciences and Engineering Research Council of Canada (NSERC) \\  
$^{b}$ &  supported by the German Federal Ministry for Education and Research (BMBF), under        
          contract numbers 05 HZ6PDA, 05 HZ6GUA, 05 HZ6VFA and 05 HZ4KHA\\                         
$^{c}$ &  supported in part by the MINERVA Gesellschaft f\"ur Forschung GmbH, the Israel Science   
          Foundation (grant no. 293/02-11.2) and the U.S.-Israel Binational Science Foundation \\  
$^{d}$ &  supported by the Israel Science Foundation\\                                             
$^{e}$ &  supported by the Italian National Institute for Nuclear Physics (INFN) \\                
$^{f}$ &  supported by the Japanese Ministry of Education, Culture, Sports, Science and Technology 
          (MEXT) and its grants for Scientific Research\\                                          
$^{g}$ &  supported by the Korean Ministry of Education and Korea Science and Engineering          
          Foundation\\                                                                             
$^{h}$ &  supported by the Netherlands Foundation for Research on Matter (FOM)\\                   
$^{i}$ &  supported by the Polish State Committee for Scientific Research, project no.             
          DESY/256/2006 - 154/DES/2006/03\\                                                        
$^{j}$ &  partially supported by the German Federal Ministry for Education and Research (BMBF)\\   
$^{k}$ &  supported by RF Presidential grant N 1456.2008.2 for the leading                         
          scientific schools and by the Russian Ministry of Education and Science through its      
          grant for Scientific Research on High Energy Physics\\                                   
$^{l}$ &  supported by the Spanish Ministry of Education and Science through funds provided by     
          CICYT\\                                                                                  
$^{m}$ &  supported by the Science and Technology Facilities Council, UK\\                         
$^{n}$ &  supported by the US Department of Energy\\                                               
$^{o}$ &  supported by the US National Science Foundation. Any opinion,                            
findings and conclusions or recommendations expressed in this material                             
are those of the authors and do not necessarily reflect the views of the                           
National Science Foundation.\\                                                                     
$^{p}$ &  supported by the Polish Ministry of Science and Higher Education                         
as a scientific project (2006-2008)\\                                                              
$^{q}$ &  supported by FNRS and its associated funds (IISN and FRIA) and by an Inter-University    
          Attraction Poles Programme subsidised by the Belgian Federal Science Policy Office\\     
$^{r}$ &  supported by an FRGS grant from the Malaysian government\\                               
\end{tabular}                                                                                      
                                                           %
                                                           %

\pagenumbering{arabic} 
\pagestyle{plain}

\section{\bf Introduction}
\label{s:intro}

Deep inelastic scattering (DIS) of leptons off nucleons has proved to
be a key process in the understanding of the structure of the proton and testing 
of the Standard Model (SM).
Neutral current (NC) DIS is mediated by photons and $Z$ bosons and is
sensitive to all quark flavours. However, at leading order only up-type quarks
and down-type antiquarks contribute to $e^-p$ charged current (CC) DIS.
Thus this process is a powerful probe of flavour-specific 
parton distribution functions (PDFs).
Due to the chiral nature of the weak interaction, the SM predicts
a linear dependence of the CC cross section on the degree of longitudinal
polarisation of the electron beam. The cross section is expected to be zero 
for a right-handed electron beam.

The HERA $ep$ collider allowed the exploration of CC DIS~\mcite{pl:b324:241,zfp:c67:565,pl:b379:319,epj:c13:609,epj:c19:269,epj:c21:33,epj:c30:1,prl:75:1006,zfp:c72:47,epj:c12:411,plb539cf,epj:c32:1} up to much higher $Q^{2}$ 
than previously possible in fixed-target
experiments~\mcite{zfp:c25:29,zfp:c49:187,zfp:c53:51,zfp:c62:575}. This paper presents measurements of 
the cross sections for $e^{-}p$ CC
DIS with longitudinally polarised electron beams. The measured
cross sections are compared to SM predictions and
previous ZEUS measurements of $e^{+}p$ CC DIS
with longitudinally polarised positron beams~\cite{pl:b637:210}.
Similar results in $e^{+}p$ CC DIS have been published by the 
H1 Collaboration~\cite{pl:b634:173}.

%
%
\section{\bf Kinematic variables and cross sections}
\label{s:Kincross}

Deep inelastic lepton-proton scattering can be described in
terms of the kinematic variables $x$, $y$ and $Q^2$.  The variable
$Q^2$ is defined as $Q^2 = -q^2 = -(k-k')^2$ where $k$ and $k'$ are
the four-momenta of the incoming and scattered lepton,
respectively. Bjorken $x$ is defined by $x=Q^2/2P \cdot q$ where $P$
is the four-momentum of the incoming proton. The variable $y$ is
defined by $y=P\cdot q/P \cdot k$. The variables $x$, $y$ and $Q^2$
are related by $Q^2=sxy$, where $s=4E_e E_p$ is the square of the
lepton-proton centre-of-mass energy (neglecting the masses of the
incoming particles) and $E_{e}$ and $E_{p}$ are the energies of the 
incoming electron and proton, respectively.

The longitudinal polarisation of the electron beam, $P_{e}$, is defined as
\begin{equation}
P_{e}=\frac{N_{R}-N_{L}}{N_{R}+N_{L}}, \nonumber
\label{eqn:pol}
\end{equation}
where $N_{R}$ and $N_{L}$ are the numbers of right- and left-handed electrons in the beam.
The electroweak Born-level cross section for the CC reaction,
$e^{-}p\rightarrow \nu_{e}X$, with longitudinally polarised electron
beams, can be expressed as~\cite{devenish:2003:dis}
\begin{equation}
\frac{d^2 \sigma ^{\rm CC} (e^-
  p)}{dxdQ^2}=(1-P_{e})\frac{G_{F}^{2}}{4\pi
  x}\bigg(\frac{M_{W}^{2}}{M_{W}^{2}+Q^{2}}\bigg) ^{2} \bigg[
  Y_{+}F_{2}^{\rm CC}(x,Q^{2})+Y_{-}xF_{3}^{\rm CC}(x,Q^{2})
  -y^{2}F_{L}(x,Q^{2})
  \bigg], \nonumber
  \label{eBorn}
\end{equation}
where $G_{F}$ is the Fermi constant, $M_{W}$ is the mass of the $W$
boson and $Y_{\pm}=1\pm(1-y)^{2}$. The longitudinal structure function gives a negligible contribution to the cross section, except at values of $y$ close to 1. 
Within the framework of the quark-parton model, 
the structure functions $F_{2}^{\rm CC}$ and $xF_{3}^{\rm CC}$ 
for $e^{-}p$ collisions can be written in terms of sums and differences of 
quark and anti-quark PDFs as follows:
\begin{equation}
F_{2}^{\rm CC} = x[u(x,Q^{2}) + c(x,Q^{2})+\bar{d}(x,Q^{2})+\bar{s}(x,Q^{2})],
 \nonumber
\end{equation}
\begin{equation}
xF_{3}^{\rm CC} = x[u(x,Q^{2}) + c(x,Q^{2})-\bar{d}(x,Q^{2})-\bar{s}(x,Q^{2})], \nonumber
\end{equation}
where, for example, the PDF $u(x,Q^{2})$ gives the number density of up quarks with momentum-fraction $x$ at a given $Q^{2}$. Since the top-quark mass is large and the off-diagonal elements of the CKM matrix are small~\cite{jp:g33:1}, the contribution from third-generation quarks may be ignored~\cite{katz:2000:hera}.

\section{\bf Experimental apparatus}
\label{s:detector}

A detailed description of the ZEUS detector can be found 
elsewhere~\cite{zeus:1993:bluebook}. A brief outline of the 
components most relevant for this analysis is given
below.

Charged particles were tracked in the central tracking detector (CTD)~\citeCTD,
which operated in a magnetic field of $1.43\Tesla$ provided by a thin 
superconducting solenoid. The CTD consisted of 72~cylindrical drift chamber 
layers, organised in nine~superlayers covering the polar-angle\ZcoosysfnB~region 
\mbox{$15^\circ<\theta<164^\circ$}. 
A silicon microvertex detector (MVD)~\cite{nim:a581:656} was installed 
between the beampipe and the inner radius of the CTD. 
The MVD was organised into a barrel with three cylindrical layers and a forward 
section with four planar layers perpendicular to the HERA beam direction.
Charged-particle tracks were reconstructed using information from the CTD and MVD. 

The high-resolution uranium--scintillator calorimeter (CAL)~\citeCAL consisted 
of three parts: the forward (FCAL), the barrel (BCAL) and the rear (RCAL)
calorimeter, covering 99.7\% of the solid angle around the nominal interaction point. 
Each part was subdivided transversely into towers and
longitudinally into one electromagnetic section (EMC) and either one (in RCAL)
or two (in BCAL and FCAL) hadronic sections (HAC). The smallest subdivision of
the calorimeter was called a cell.  The CAL relative energy resolutions, 
as measured under
test-beam conditions, were $\sigma(E)/E=0.18/\sqrt{E}$ for electrons and
$\sigma(E)/E=0.35/\sqrt{E}$ for hadrons, with $E$ in $\Gev$. The timing resolution of the CAL was better than 1~ns 
for energy deposits exceeding 4.5~\gev.

An iron structure that surrounded the CAL was instrumented as a backing
calorimeter (BAC)~\cite{nim:a313:126} to measure energy leakage from the CAL. Muon chambers in the forward, 
barrel and rear~\cite{nim:a333:342} regions were used in this analysis to veto background events induced by 
cosmic-ray or beam-halo muons.

The luminosity was measured using the Bethe-Heitler reaction $ep
\rightarrow e \gamma p$ with the luminosity detector which consisted of
two independent systems, a photon calorimeter and a magnetic
specrometer.

The lepton beam in HERA became naturally transversely polarised through the Sokolov-Ternov effect~\cite{sovpdo:8:1203,sovjnp:b9:238}.
The characteristic build-up time for the HERA accelerator was approximately 40~minutes. 
Spin rotators on either side of the ZEUS detector changed the 
transverse polarisation of the beam into longitudinal polarisation and back again. The electron beam polarisation was measured using 
two independent polarimeters, the transverse polarimeter (TPOL)~\cite{nim:a329:79} and the longitudinal polarimeter (LPOL)~\cite{nim:a479:334}. 
Both devices exploited the spin-dependent cross section for Compton scattering of circularly polarised photons off electrons to measure the beam polarisation.
The luminosity and polarisation measurements were made over times that
were much shorter than the polarisation build-up time.

\begin{figure}
  \begin{center}
    \includegraphics*[width=.8\textwidth]{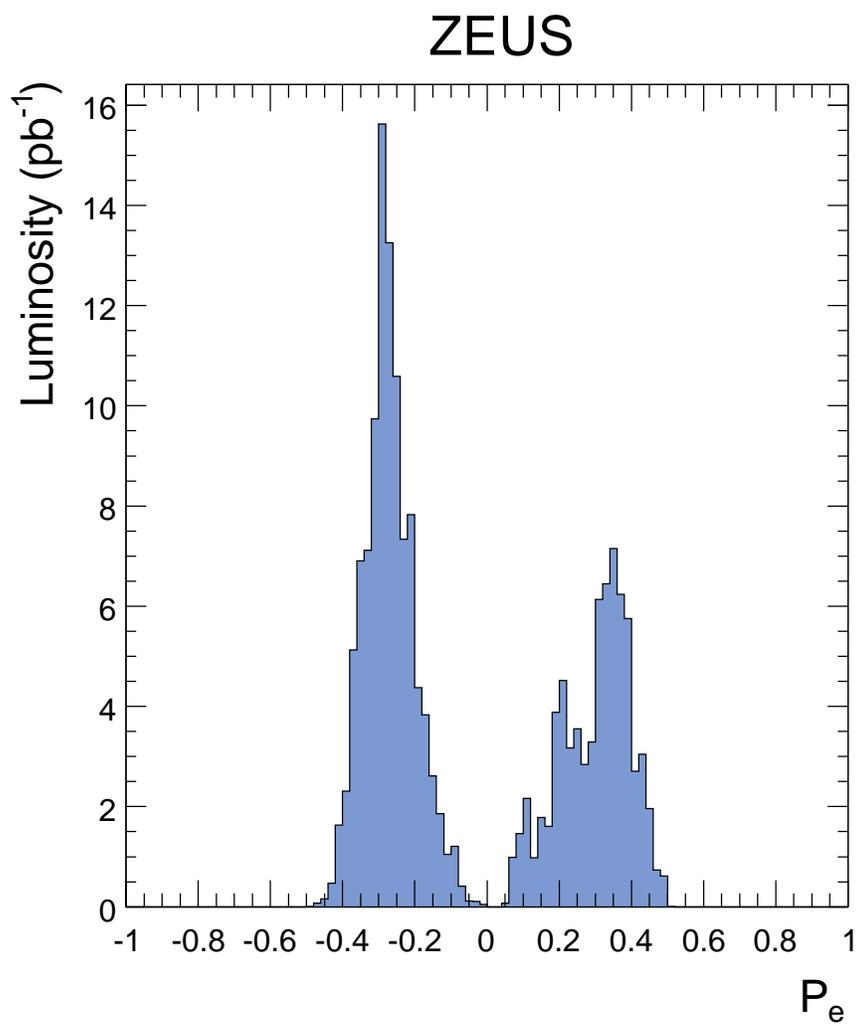}
 \end{center}
  \caption{
    The integrated luminosity collected as a function of 
    the longitudinal polarisation of the electron beam.
    }
  \label{f:lumipol}
\end{figure}

The measurements are based on data samples collected with the ZEUS
detector from 2004 to 2006 when HERA collided protons of
energy $920\gev$ with electrons of energy $27.5\gev$, yielding
collisions at a centre-of-mass energy of $318\gev$. 
The integrated luminosities of the data
samples were $104\pb^{-1}$ and $71\pb^{-1}$ at mean luminosity weighted 
polarisations of $-0.27$ and $+0.30$, respectively.
Figure~\ref{f:lumipol} shows the luminosity collected as a function of the 
longitudinal polarisation of the electron beam.

\section{Monte Carlo simulation}
\label{s:MCsimulation}
Monte Carlo (MC) simulations were used to determine the efficiency for 
selecting events and the accuracy of kinematic 
reconstruction, to estimate the background rates from $ep$ processes other than CC DIS and to extract
cross sections for the full kinematic region.
A sufficient number of
events was generated to ensure that the statistical uncertainties arising from the MC simulation were negligible compared to those of the data.
The MC samples were normalised to the total
integrated luminosity of the data.

Charged current DIS events, including electroweak radiative effects, were 
simulated using the {\sc heracles} 4.6.3~\cite{cpc:69:155,*spi:www:heracles} 
program with the
{\sc djangoh} 1.3~\cite{spi:www:djangoh11} interface to the MC generators that provide the hadronisation. Initial-state radiation, vertex and 
propagator corrections and two-boson exchange are included in {\sc heracles}.
The parameters of the SM were set to the PDG~\cite{jp:g33:1} values.
The events were generated using the CTEQ5D~\cite{epj:c12:375} PDFs.
The colour-dipole model of {\sc ariadne} 4.10~\cite{cpc:71:15}
was used to simulate $\mathcal{O}(\alpha_{S})$ plus leading-logarithmic corrections to the result of the quark-parton model. This program uses the Lund string model of {\sc jetset} 7.4~\cite{cpc:39:347,*cpc:43:367,*cpc:82:74} for the hadronisation.
A set of NC DIS events generated with {\sc djangoh}
 was used to estimate the NC contamination in the CC sample.
Photoproduction background was estimated using events 
simulated with {\sc herwig} 5.9~\cite{cpc:67:465}.
The background from 
$W$ production was estimated
using the {\sc epvec} 1.0~\cite{np:b375:3} generator, and the background from
production of charged-lepton pairs was generated with the {\sc grape} 1.1~\cite{cpc:136:126} program.

The vertex distribution in data is a crucial input to the MC simulation for the
correct evaluation of the event-selection efficiency. Therefore, the $Z$-vertex
distribution used in the MC simulation was determined from a sample of NC DIS
events in which the event-selection efficiency was independent of $Z$.

The ZEUS detector response was simulated with a program based on 
{\sc geant}~3.21~\cite{tech:cern-dd-ee-84-1}.  The simulated events were subjected 
to the same trigger requirements as 
the data, and processed by the same reconstruction programs.

\section{Reconstruction of kinematic variables}
\label{s:reconstruction}
The principal signature of CC DIS at HERA
is large missing transverse momentum, $\PTM$,
arising from the energetic final-state neutrino which escapes detection. 
$\PTM$ is related to the total hadronic momentum, ${P}_{T}$, by $\PTM^{2}  =  (-\overrightarrow{P}_{T})^{2}$, where
\begin{equation}  
  (\overrightarrow{P}_{T})^2 = 
  \left( \sum\limits_{i} E_i \sin \theta_i \cos \phi_i \right)^2
+ \left( \sum\limits_{i} E_i \sin \theta_i \sin \phi_i \right)^2. \nonumber
  \label{eq:pt}
\end{equation}
The sums run over all CAL energy deposits, $E_i$ and
$\theta_i$ and $\phi_i$ are the polar and azimuthal angles.
The calorimeter energy deposits are clustered cell energies corrected for energy loss 
in inactive material and reconstruction deficiencies~\cite{epj:c11:427}.
The polar angle of the hadronic system, $\gamma_h$, is defined by
\begin{equation}
\cos\gamma_h = ( (\overrightarrow{P}_{T})^2 - \delta^2)/( (\overrightarrow{P}_{T})^2 + \delta^2), \nonumber
\end{equation}
where
$\delta = \sum\limits_{i} E_i ( 1 - \cos \theta_{i} ) 
= \sum\limits_{i} (E-P_Z)_{i}$.
In the naive quark-parton model,
$\gamma_h$ is the angle of the scattered quark.
Finally, the total transverse energy,
$E_T$, is given by
$E_T    = \sum\limits_{i} E_i \sin \theta_i$. 

The ratio of the parallel, $V_{P}$, and anti-parallel, $V_{AP}$, components of the
hadronic transverse momentum can be used to distinguish CC DIS from photoproduction events. 
These variables are defined as
\begin{equation}
V_{P} = \sum\limits_{i} \overrightarrow{P}_{T,i} \cdot \overrightarrow{n}_{P_{T}}~~~{\rm for}~~\overrightarrow{P}_{T,i} \cdot \overrightarrow{n}_{P_{T}}>0,
\nonumber
\end{equation}
\begin{equation}
V_{AP} = -\sum\limits_{i} \overrightarrow{P}_{T,i} \cdot \overrightarrow{n}_{P_{T}}~~~{\rm for}~~\overrightarrow{P}_{T,i} \cdot \overrightarrow{n}_{P_{T}}<0,
\nonumber
\end{equation}
where the sums are performed over all calorimeter cells and $\overrightarrow{n}_{P_{T}} = \overrightarrow{P}_{T}/P_{T}$.

 The kinematic variables were reconstructed
using the Jacquet-Blondel method \cite{proc:epfacility:1979:391}.
The estimators of $y$, $Q^2$ and $x$ are:
$y_{\rm{JB}} = \delta/(2E_e)$,
$Q^2_{\rm{JB}} = P_{T}^2/(1-y_{\rm{JB}})$, and
$x_{\rm{JB}} = Q^2_{\rm{JB}}/(sy_{\rm{JB}})$.

The resolution in $Q^2$ 
is about $20\%$. The resolution in $x$ improves from about $20$\%
at $x=0.01$ to about $5\%$ at $x=0.5$.
The resolution in $y$ ranges from about $14\%$ at $y=0.05$ to about 
$8\%$ at $y=0.83$.

\section{Event selection}
\label{s:EvSel}
Charged current DIS candidates were selected by requiring a large $\PTM$.
The main sources of background came from NC DIS and high-$E_T$
photoproduction in which the finite energy resolution of the CAL
or energy that escapes detection
can lead to significant measured missing transverse momentum.  
Non-$ep$ events such as beam-gas interactions,
beam-halo muons or cosmic rays can also cause substantial 
imbalance in the measured transverse
momentum and constitute additional sources of background.
The selection criteria described below were imposed
to separate CC events from all backgrounds.

\subsection{Trigger selection}
\label{ss:Trigger}
ZEUS had a three-level trigger system~\cite{zeus:1993:bluebook,uproc:chep:1992:222,nim:a580:1257}. 
At the first level, only coarse calorimeter and tracking information was available.
Events were selected using criteria based on the energy, transverse 
energy and missing transverse momentum measured in the calorimeter. 
Generally, events were triggered with low thresholds on these quantities if a
coincidence with CTD tracks from the event vertex occurred, while higher 
thresholds were required for events with no CTD 
tracks.

At the second level, timing information from the calorimeter was used
to reject events inconsistent with the bunch-crossing time. 
In addition, the topology of the CAL energy deposits was used to reject background events. 
In particular, a tighter cut was made on missing transverse momentum, since the resolution in this variable was
better at the second level than at the first level.

At the third level, full track reconstruction and vertex finding were performed
and used to reject candidate events with a vertex inconsistent with an $ep$
interaction. Cuts were applied to calorimeter quantities and reconstructed
tracks to further reduce beam-gas contamination.

\subsection{Offline selection}

When $\gamma_h$ is large,
charged-particle tracks can be used to reconstruct 
the event vertex, strongly suppressing non-$ep$ backgrounds.
For CC events with small $\gamma_h$, the charged particles of
the hadronic final state 
are often outside the acceptance of the tracking detectors.
Such events populate the high-$x$ region of the kinematic plane.
The events were classified according to $\gamma_0$, the value of
$\gamma_h$ measured with respect to the nominal interaction point.
For events with large $\gamma_0$ the kinematic quantities were recalculated using the
$Z$-coordinate of the event vertex ($Z_{\rm vtx}$) determined
from charged-particle tracks.

In events with $\gamma_{0}>0.4$ rad a reconstructed vertex was required. 
Additional requirements for event selection are given below.

\begin{itemize}
\item{selection of CC events: 
    \begin{itemize}
    \item{$| Z_{\rm vtx} | < 50$~cm;} 
    \item{$\PTM > 12~\gev$;}
    \end{itemize}}
\item{rejection of beam-gas events: 
    \begin{itemize}
    \item {$\PTM ' > 10\gev$ and $\PTM '' > 8\gev$ where $\PTM '$ is the missing
        transverse momentum calculated excluding the ring of FCAL towers 
        closest to the beam pipe and $\PTM ''$ is the corresponding quantity calculated 
        excluding the two rings of FCAL towers closest to the beam pipe.
        These requirements strongly suppress beam-gas events while
        maintaining high efficiency for CC events;}
    \item {tracks associated with the event vertex with
        transverse momentum in excess of $0.2\gev$ and a polar
        angle in the range $15^\circ$ to $164^\circ$ were defined as ``good''
        tracks. In order to remove
        beam-gas background, at least one such track was required and a cut was 
        also applied in two dimensions on the
        number of good tracks versus the total number of tracks. 
        This cut was $N_{\rm GoodTrks}>0.3\dot (N_{\rm Trks}-20)$;}
    \end{itemize}}
  \item{rejection of photoproduction:
      \begin{itemize} 
      \item{$V_{AP}/V_P <0.4$ was required for events
          with $\PTM < 30\gev$. For events with $\PTM < 20\gev$ this
          cut was tightened to $V_{AP}/V_P < 0.23$. 
          This selected events with a collimated energy flow,
          as expected from a single scattered quark;}
      \item{for charged current events, there is a correlation between the direction
        of the $\PTM$ vector calculated using tracks and that obtained using
        the CAL. The difference between these quantities was required to be less
        than 0.5~radians for \mbox{$\PTM <45\gev$}. As less background
        is expected for high $\PTM$ this requirement was loosened to less than 1.0~radian 
        for \mbox{$\PTM \geq 45\gev$};}
    \end{itemize}}
\item{rejection of NC DIS: NC DIS events in which the energies of the scattered electron or
    the jet are poorly measured can have a considerable apparent
    missing transverse momentum. To identify such events, a search for
    candidate electrons was made using isolated electromagnetic clusters in
    the CAL~\cite{nim:a365:508,*nim:a391:360} for events with $\PTM <
    30\gev$. Candidate electron clusters within the tracking acceptance were
    required to have an energy above $4\gev$ and a matching track. 
    Clusters with $\theta > 164^\circ$ were required to have 
    a transverse momentum exceeding
    $2\gev$. Events with a candidate electron satisfying the above criteria
    and $\delta > 30\gev$ were rejected, since for fully contained NC
    events, $\delta$ peaks at $2E_e = 55\gev$;}
\item{rejection of non-$ep$ background: muon-finding
    algorithms based on CAL energy deposits or
    muon-chamber signals were used to reject events produced by cosmic rays
    or muons in the beam halo.}
\end{itemize}

In events with $\gamma_{0}<0.4$ rad some requirements were tightened
to compensate for the relaxation of the track requirements. 
Additional requirements for event selection are given below.

\begin{itemize}
\item{missing transverse momentum: events were required to satisfy 
    $\PTM > 14\gev$ and $\PTM ' > 12\gev$;}
\item{rejection of non-$ep$ background: A class of background events arose
    from beam-halo muons that produced a shower inside the FCAL. To reduce
    this background, in addition to the muon-rejection cuts described  
    above, topological cuts on the transverse and longitudinal
    shower shape were imposed. These cuts rejected events in which the energy
    deposits were more collimated than for typical hadronic jets.}
\end{itemize}

The kinematic region was restricted to $Q^2_{\rm{JB}}>200\gev^2$
and $y_{\rm{JB}}<0.9$ to ensure good resolution.

\begin{figure}
  \begin{center}
    \includegraphics*[width=.8\textwidth]{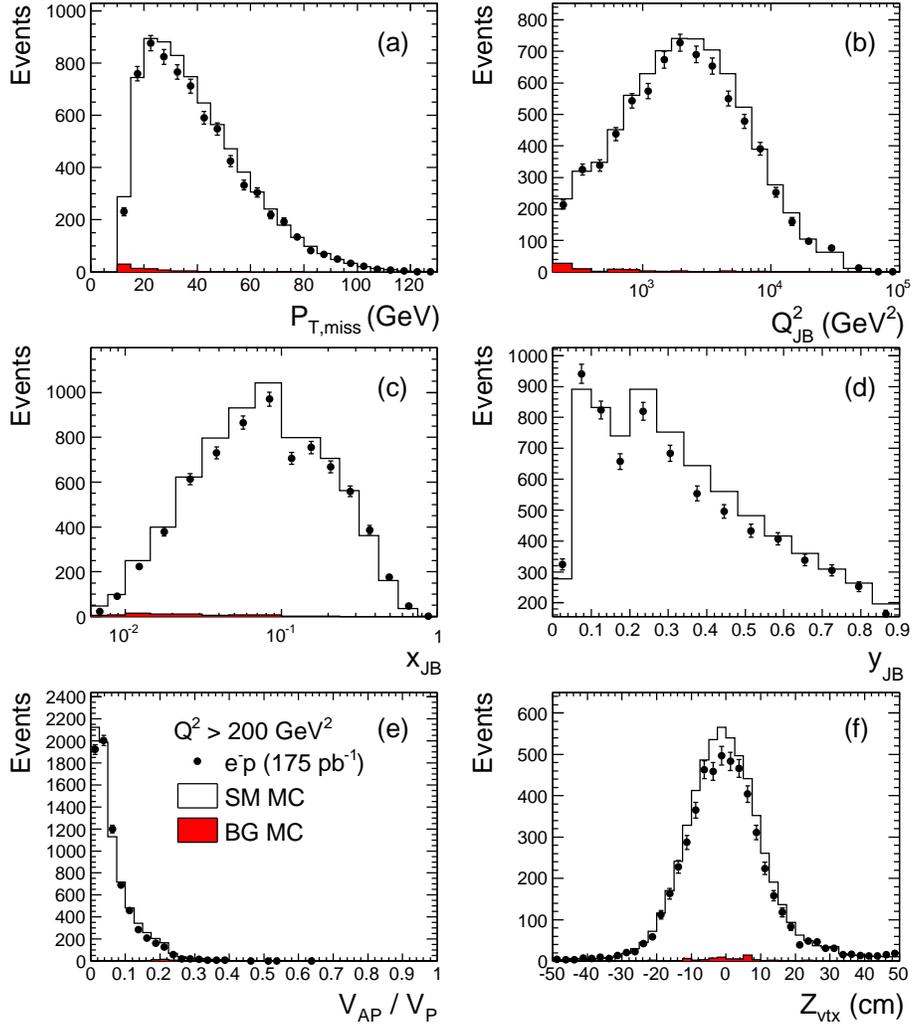}
 \end{center}
  \caption{
    Comparison of the $e^- p$ CC data sample with the expectations of
    the MC simulation as described in Section~\ref{s:MCsimulation} of the text. 
    The distributions of (a) $\PTM$, (b) $Q^{2}_{\rm JB}$,
    (c) $x_{\rm JB}$, (d) $y_{\rm JB}$, (e) $V_{AP}/V_{P}$ and (f) $Z_{\rm vtx}$ are shown.
    }
  \label{f:control}
\end{figure}

A total of 7198 events satisfied these criteria. A background contamination from $ep$ processes of 0.5\%,
 dominated by the photoproduction component, is predicted. Figure~\ref{f:control} 
compares the distributions of data events entering the final CC sample with the MC expectation 
for the sum of the CC signal and $ep$ background events. The MC simulations give a reasonable 
description of the data.
\section{Cross-section determination and systematic \\ uncertainties}

\label{s:xsect}
The measured cross section
in a particular kinematic bin, for example for $d^2\sigma/dxdQ^2$, was determined from

\begin{equation}
  \frac{d^2\sigma}{dxdQ^2} = \frac{N_{\rm{data}}-N_{\rm{bg}}}{N_{\rm{MC}}} \cdot \frac{d^2 \sigma^{\rm{SM}}_{\rm{Born}}}{dxdQ^2}, \nonumber
\end{equation}

where $N_{\rm{data}}$ is the number of data events, $N_{\rm{bg}}$ is
the number of background events estimated from the MC simulation and
$N_{\rm{MC}}$ is the number of signal MC events. The cross-section 
$\frac{d^2 \sigma^{\rm{SM}}_{\rm{Born}}}{dxdQ^2}$ is the Standard
Model prediction evaluated in the on-shell scheme~\cite{epj:c15:1}
using the PDG values for the electroweak parameters and the CTEQ5D PDFs~\cite{epj:c12:375}.
 A similar procedure was used for $d\sigma/dQ^2$, $d\sigma/dx$ and $d\sigma/dy$. 
Consequently, the acceptance, as well as the bin-centring and 
radiative corrections were all taken from the MC simulation. 
The cross-sections $ d \sigma/ d  Q^2$ and $ d \sigma/ d  x$ were 
extrapolated to the full $y$ range using the SM predictions calculated with the CTEQ5D PDFs. 

\label{s:systerr}
The systematic uncertainties in the measured cross sections were determined by
changing the analysis procedure in turn and repeating the
extraction of the cross sections. 

\begin{itemize}
\item{calorimeter energy scale: the relative uncertainty of the hadronic energy scale was 2\%. 
  Varying the energy scale of the calorimeter by this 
  amount in the detector simulation induces small shifts of the Jacquet-Blondel 
  estimators of the kinematic variables. The variation of the energy scale for each 
  of the calorimeters simultaneously up or down
  by this amount gave the systematic uncertainty 
  on the total measured energy in the calorimeter.
  The resulting systematic shifts in the measured cross sections were typically within 
  $\pm 5\%$, but increased to $\pm (20-30)\%$ in the highest $Q^{2}$ and $x$ bins of the 
  single-differential cross sections and reached $\pm 45 \%$
  in the double-differential cross section;}
\item{reconstruction: an alternative analysis~\cite{thesis:sutiak:2006} was performed using 
    jets to reconstruct the kinematic quantities and reject background. 
    The difference between the nominal and jet analyses was 
    taken as an estimate of the systematic uncertainity on the reconstruction and background rejection. 
    The difference was found to be typically within $\pm 10\%$, but increased up to $\pm (20-25)\%$ 
    in the highest $Q^{2}$ and $x$ bins of the cross sections;}
\item{background subtraction: the uncertainty in the small contribution from 
    photoproduction was estimated by varying the normalisation by $\pm
    60\%$, resulting~\cite{thesis:fry:2007, thesis:kaji:2006} in modifications of the cross sections within $\pm 2\%$;}
\item{selection criteria: in order to estimate the bias introduced into the 
    measurements from an imperfect description of the data by the MC simulation, the efficiencies 
    for each of the selection criteria were measured using the hadronic final state in NC DIS data.
    Using the measured efficiencies to extract the cross sections instead of the CC MC gave 
    changes in the cross sections that were typically within $\pm 2 \%$, except for the two-dimensional
    tracking cut which gave an effect of $10 \%$ at high $Q^{2}$;}
\item{the uncertainties associated with the trigger, choice of PDFs in the MC and the measurement of the vertex 
    positions were negligible.}
\end{itemize}

The individual uncertainties were added in quadrature 
separately for the positive and negative deviations from the
nominal cross-section values to obtain the total systematic uncertainties. 
The $\mathcal{O}(\alpha)$ electroweak corrections to CC DIS 
have been discussed by several authors~\cite{jp:g25:1387,proc:mc:1998:530}. 
Various theoretical approximations and computer codes gave differences in 
the CC cross sections of typically $\pm(1-2)\%$ or less. 
 However, the differences can be as large as ~$\pm(3-8)\%$ at high $x$ and high $y$. No uncertainty was included in the measured cross sections from this source. 

The relative uncertainty in the measured polarisation was $3.6\%$ using the LPOL and $4.2\%$ 
using the TPOL. The choice of polarimeter measurement was made on a run-by-run basis. 
The LPOL measurement was used when available, otherwise the TPOL measurement was used. The
uncertainty of $2.6\%$ on the measured total luminosity was not
included in the differential cross-section figures or the tables.

\section{Results}
\label{ss:Results}

The total cross section, corrected to the Born level of the electroweak interaction, for $e^- p$ CC DIS in the kinematic region $Q^{2}>200 \gev^{2}$ was measured to be 
\begin{equation}
\sigma^{\rm CC}(P_{e}=+0.30 \pm 0.01)=47.1 \pm 1.1 ({\rm stat.}) \pm 2.2 ({\rm syst.})~{\rm pb}, \nonumber
\end{equation}
\begin{equation}
\sigma^{\rm CC}(P_{e}=-0.27 \pm 0.01)=83.1 \pm 1.2 ({\rm stat.}) \pm 3.3 ({\rm syst.})~{\rm pb}. \nonumber 
\end{equation}

\begin{figure}
  \begin{center}
    \includegraphics[width=.8\textwidth]{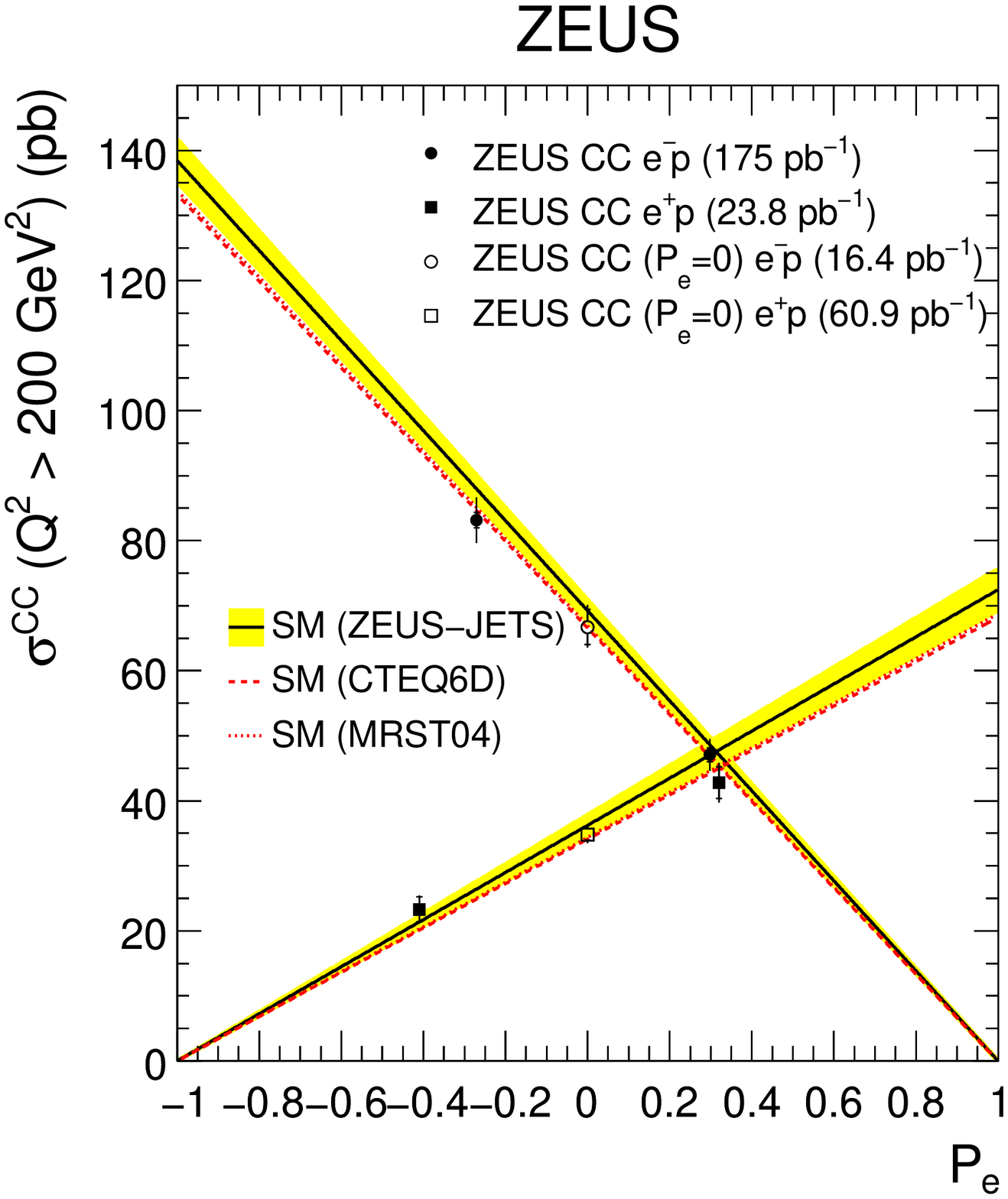}
  \end{center}
  \caption{ 
    The total cross sections for $e^- p$ and $e^+ p$ CC DIS as a function of the longitudinal 
    polarisation of the lepton beam. The lines show the predictions of the SM evaluated using
    the ZEUS-JETS, CTEQ6D and MRST04 PDFs. The shaded bands show the
    experimental uncertainty from the ZEUS-JETS fit.
    }
  \label{f:cctotall}

\end{figure}

\begin{figure}
  \begin{center}
    \includegraphics[width=.8\textwidth]{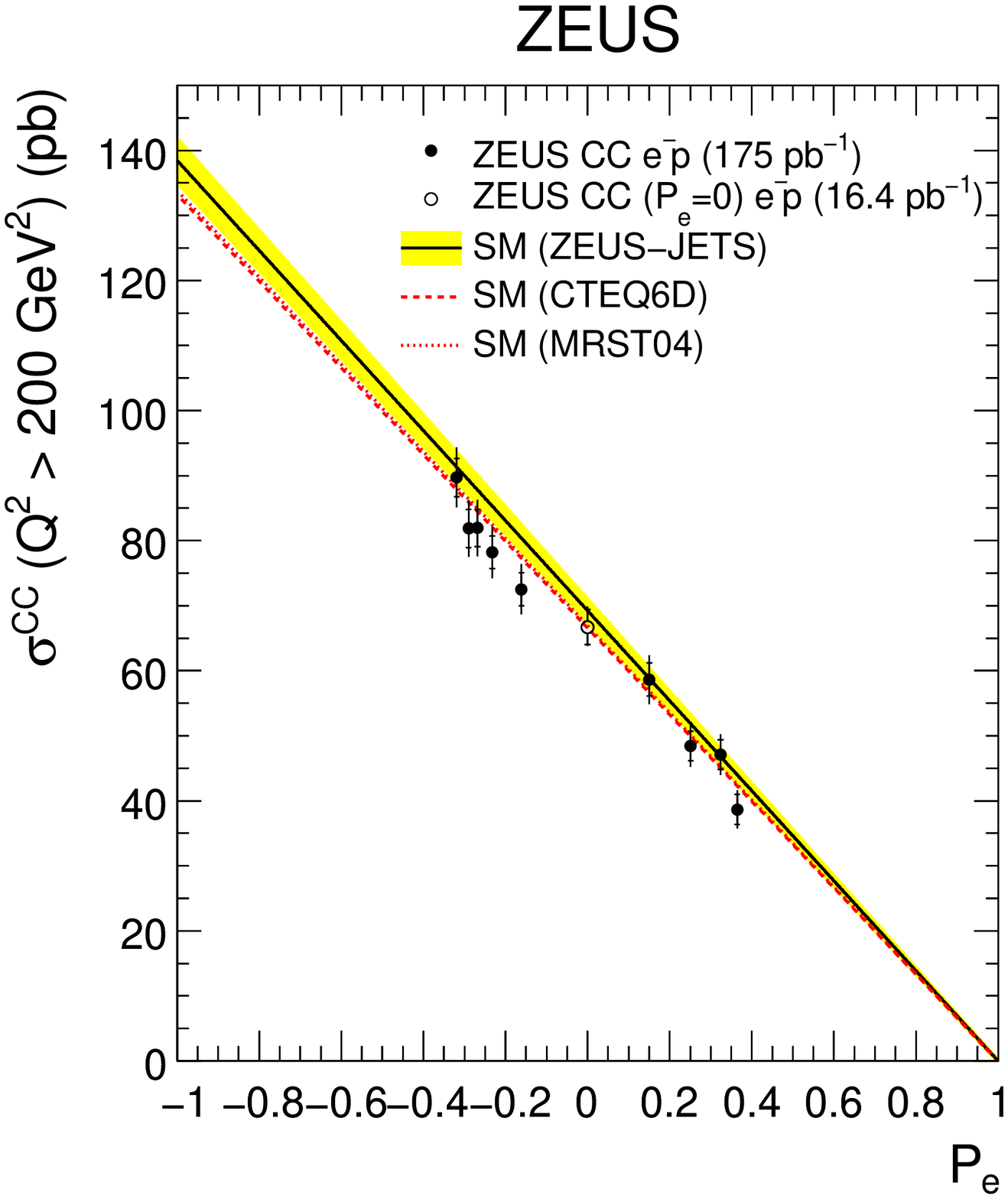}
  \end{center}
  \caption{ 
    The total cross sections for $e^- p$ CC DIS as a function of the longitudinal 
    polarisation of the electron beam. The lines show the predictions of the SM evaluated using
    the ZEUS-JETS, CTEQ6D and MRST04 PDFs. The shaded band shows the
    experimental uncertainty from the ZEUS-JETS fit.
    }
  \label{f:cctotmany}

\end{figure}

The uncertainty in the measured luminosity is included in the systematic 
uncertainty. The total cross section is shown as a function of the longitudinal polarisation 
of the lepton beam in Fig.~\ref{f:cctotall}, including previous ZEUS measurements 
from both $e^{-}p$ and $e^{+}p$ data~\cite{plb539cf,epj:c32:1,pl:b637:210}. Figure~\ref{f:cctotmany} 
shows only the $e^{-}p$ data, with a finer binning to 
emphasise the dependence on the lepton beam polarisation. The
cross-section values are tabulated in Table~\ref{t:manybins}. 
The data are compared to the SM predictions evaluated at next-to-leading-order in QCD using the 
ZEUS-JETS~\cite{epj:c42:1}, CTEQ6D~\cite{jhep:0207:012} and MRST04~\cite{pl:b604:61} PDFs which describe the data well.

\begin{figure}
  \begin{center}
    \includegraphics[width=.8\textwidth]{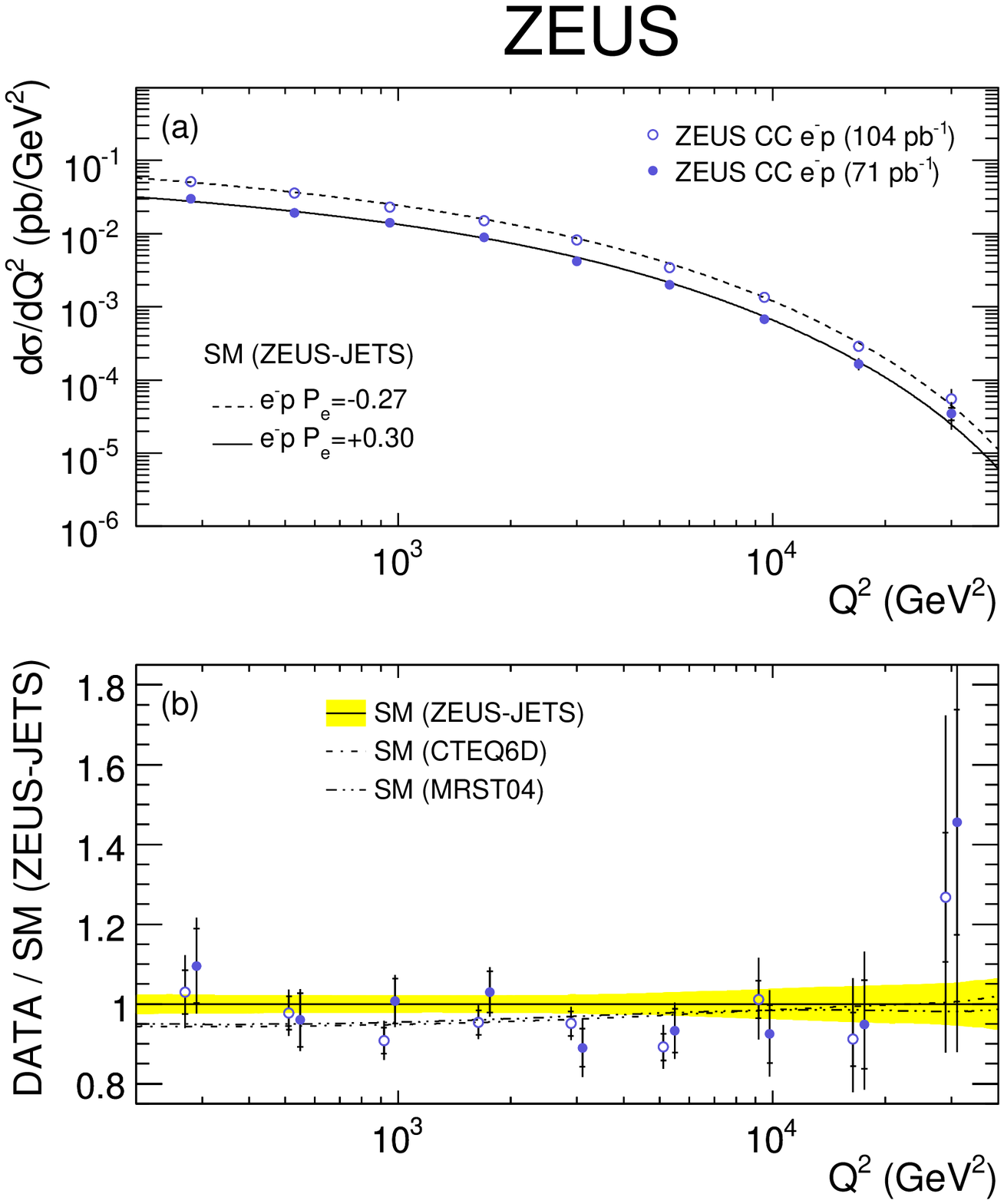}
  \end{center}
  \caption{ 
    (a) The $e^-p$ CC DIS cross-section $d\sigma/dQ^2$ for data
    and the Standard Model expectation evaluated using
    the ZEUS-JETS PDFs.
    The postive (negative) polarisation data are shown as the filled (open) points, the statistical uncertanties 
    are indicated by the inner error bars (delimited by horizontal lines) 
    and the full error bars show the total uncertainty obtained by adding 
    the statistical and systematic contributions in quadrature.
    (b) The ratio of the measured cross section, $d\sigma/dQ^2$, to the
    Standard Model expectation evaluated using the ZEUS-JETS fit.
    The shaded band shows the experimental uncertainty from the ZEUS-JETS fit.}
  \label{f:dsdq2}
\end{figure}

\begin{figure}
  \begin{center}
    \includegraphics[width=.8\textwidth]{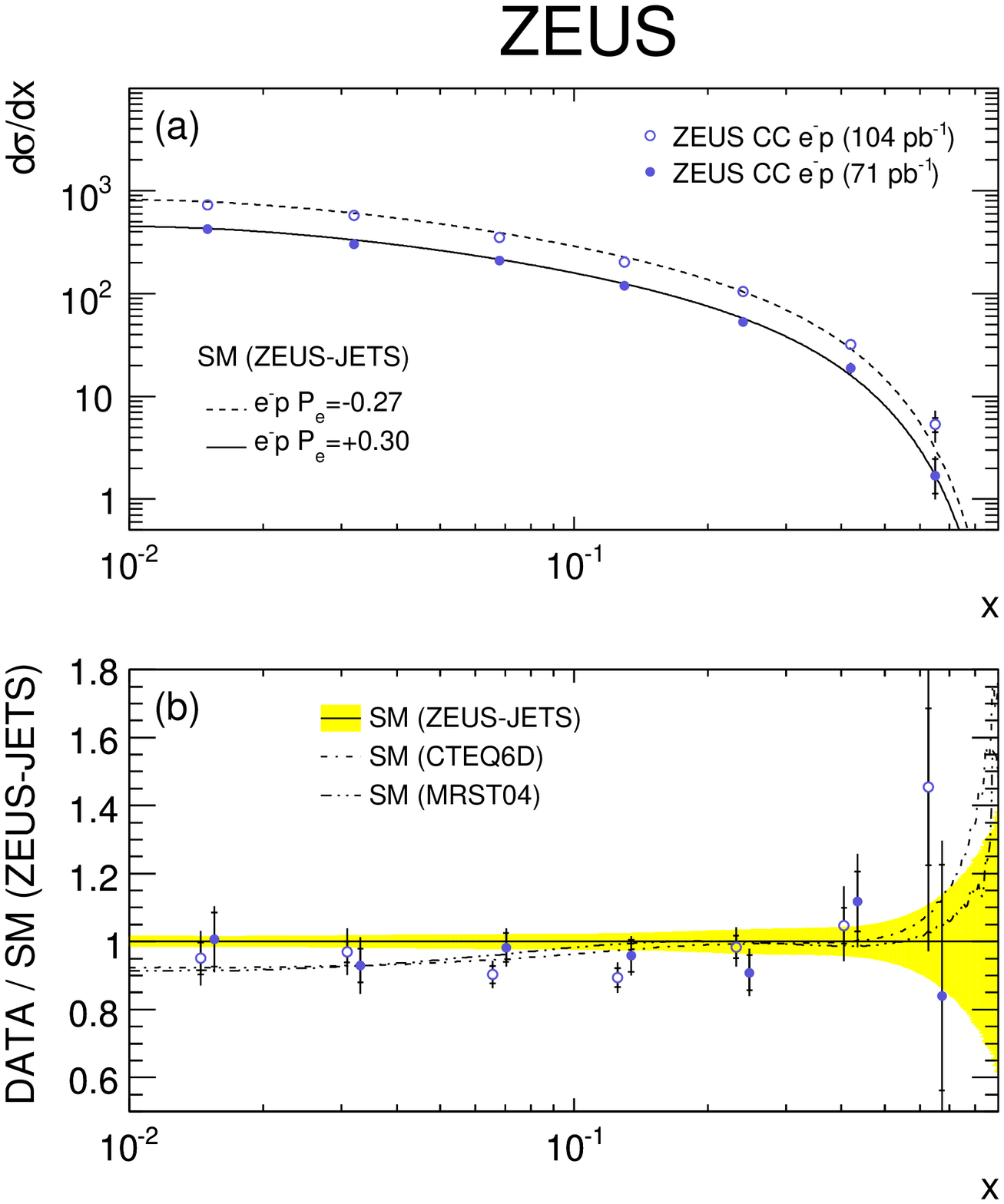}
  \end{center}
  \caption{ 
    (a) The $e^-p$ CC DIS cross-section $d\sigma/dx$ for data
    and the Standard Model expectation evaluated using
    the ZEUS-JETS PDFs.
    The postive (negative) polarisation data are shown as the filled (open) points, the statistical uncertanties
    are indicated by the inner error bars (delimited by horizontal lines) 
    and the full error bars show the total uncertainty obtained by adding 
    the statistical and systematic contributions in quadrature.
    (b) The ratio of the measured cross section, $d\sigma/dx$, to the
    Standard Model expectation evaluated using the ZEUS-JETS fit.
    The shaded band shows the experimental uncertainty from the ZEUS-JETS fit.}
  \label{f:dsdx}
\end{figure}

\begin{figure}
  \begin{center}
    \includegraphics[width=.8\textwidth]{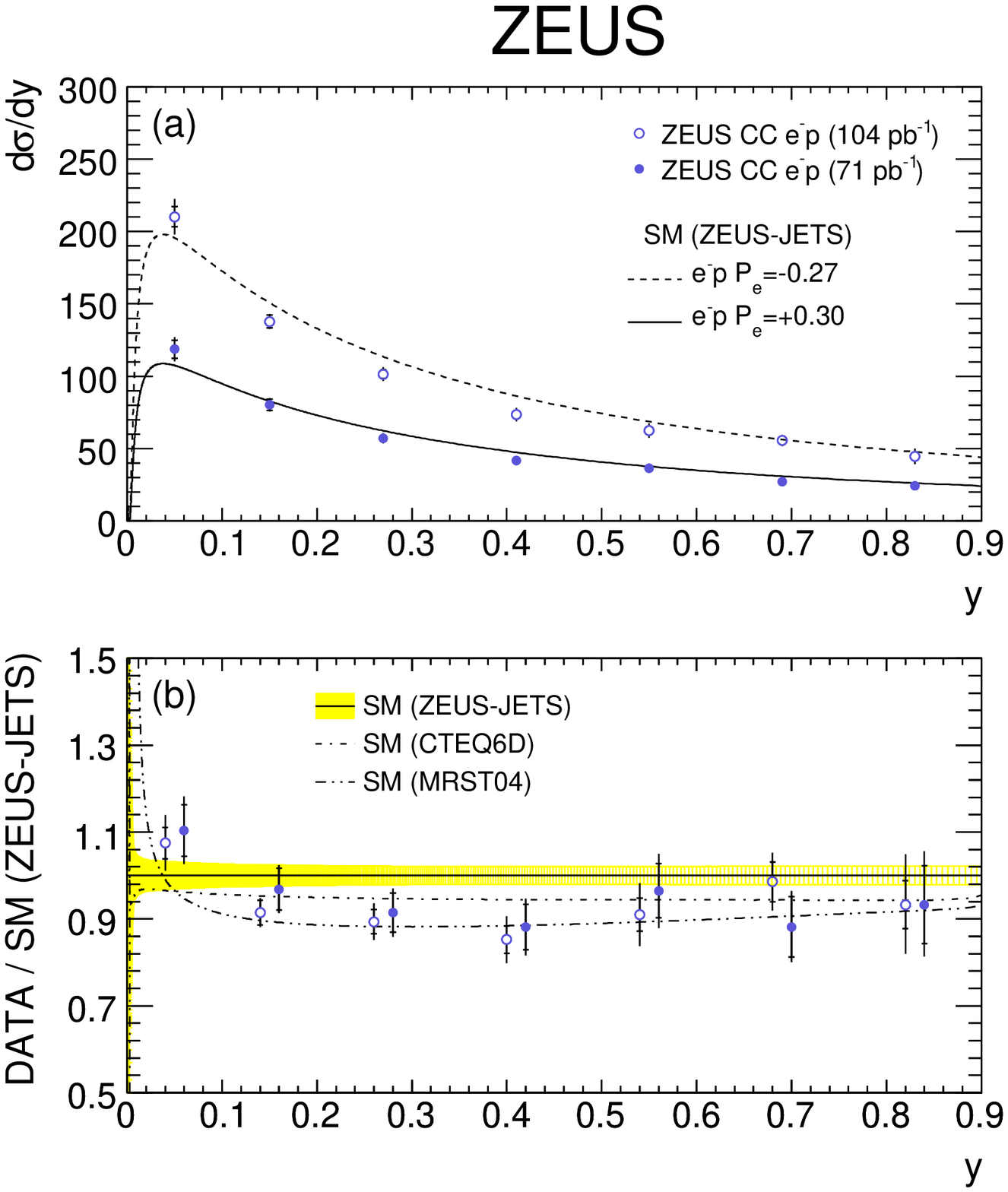}
  \end{center}
  \caption{ 
    (a) The $e^-p$ CC DIS cross-section $d\sigma/dy$ for data
    and the Standard Model expectation evaluated using
    the ZEUS-JETS PDFs.
    The postive (negative) polarisation data are shown as the filled (open) points, the statistical uncertanties
    are indicated by the inner error bars (delimited by horizontal lines) 
    and the full error bars show the total uncertainty obtained by adding 
    the statistical and systematic contributions in quadrature.
    (b) The ratio of the measured cross section, $d\sigma/dy$, to the
    Standard Model expectation evaluated using the ZEUS-JETS fit.
    The shaded band shows the experimental uncertainty from the ZEUS-JETS fit.}
  \label{f:dsdy}
\end{figure}

The single-differential cross-sections $d\sigma/dQ^{2}$, $d\sigma/dx$ and $d\sigma/dy$ 
for CC DIS are shown in Figs.~\ref{f:dsdq2},~\ref{f:dsdx}
and~\ref{f:dsdy} and given in Tables~\ref{t:single-diff}, \ref{t:uncorr_single_pos} and \ref{t:uncorr_single_neg}. 
The measurements for positive and negative longitudinal polarisation differ by a constant factor which is independent of the kinematic variables. The effects are well described by the SM evaluated using the
ZEUS-JETS, CTEQ6D and MRST04 PDFs. The precision of the data is
comparable to the uncertainties in the SM predictions, therefore these
data have the potential to further constrain the PDFs.

\begin{figure}
  \begin{center}
    \includegraphics[width=.8\textwidth]{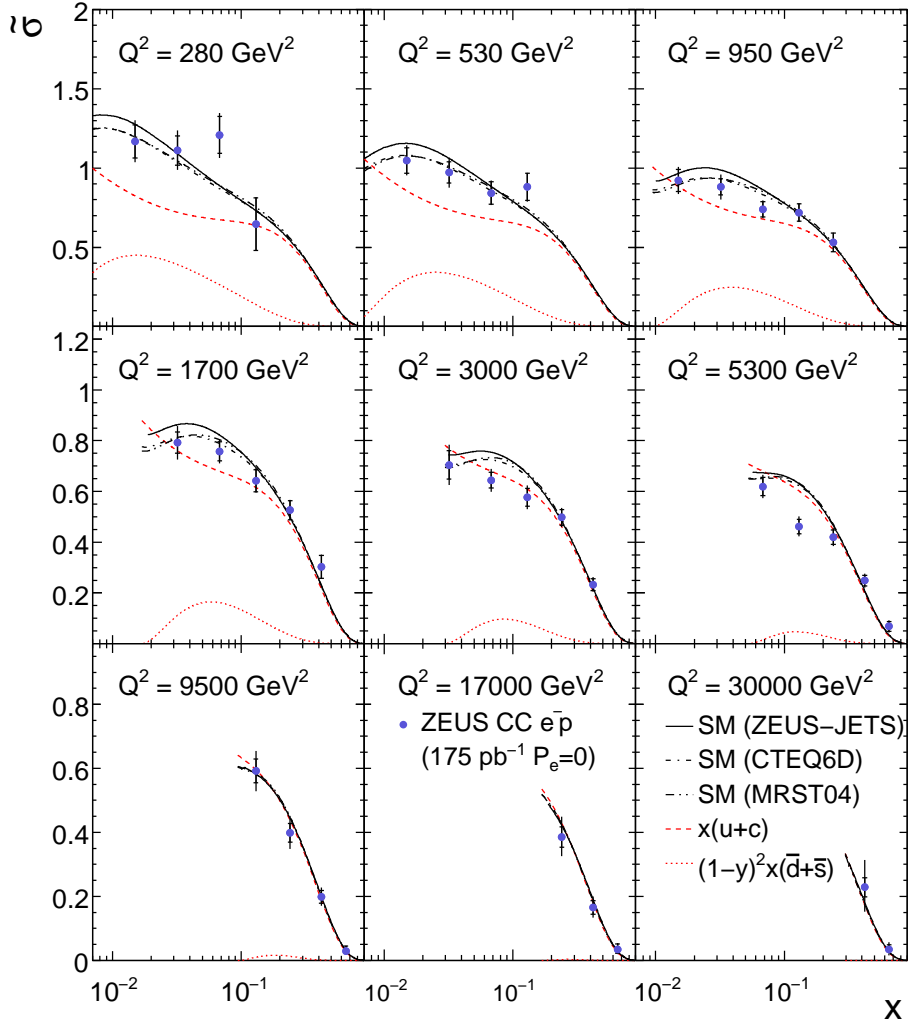}
  \end{center}
  \caption{ 
    The $e^- p$ CC DIS reduced cross section plotted as a function of $x$ for fixed $Q^{2}$.  
    The circles represent the data points
    and the curves show the predictions of the SM evaluated using
    the ZEUS-JETS, CTEQ6D and MRST04 PDFs.  
    The dashed and dotted lines show the contributions of the PDF 
    combinations $x(u+c)$ and $(1-y)^{2}x(\bar{d}+\bar{s})$, respectively.
    }
  \label{f:double}
\end{figure}

The reduced double-differential cross section, $\tilde{\sigma}$, is defined as
\begin{equation}
   \tilde{\sigma} = \left[\frac{G^2_F}{2 \pi x } \Biggl( \frac{M^2_W}{M^2_W + Q^2}\Biggr)^2 \right]^{-1}
{\frac{d^2\sigma}{dx \, dQ^2}}. \nonumber
\end{equation}
At leading order in QCD, $\tilde{\sigma}({e^- p \rightarrow \nu_e X})$
depends on the quark momentum distributions as follows:
\begin{equation}
  \tilde{\sigma} (e^- p \rightarrow {\nu}_e X) = x\left[u + c + (1-y)^2 (\bar{d} + \bar{s}) \right]. \nonumber
\end{equation}
The reduced cross section was measured in the kinematic range \mbox{$280<Q^2<30\,000 \gev^2$} and \mbox{$0.015<x<0.65$} and 
is shown as a function of $x$, at fixed values of $Q^{2}$ in
Fig.~\ref{f:double} and tabulated in Tables~\ref{t:double}, \ref{t:uncorr_double_pos} and \ref{t:uncorr_double_neg}. 
The data points were corrected to $P_{e}=0$ using the SM prediction.
The predictions of the SM evaluated using the ZEUS-JETS, CTEQ6D and MRST04 PDFs give a good description of the data. 
The contributions from the PDF combinations $(u + c)$ and $(\bar{d} + \bar{s})$, obtained 
in the \MSbar scheme from the ZEUS-JETS fit, are shown separately. 

\begin{figure}
  \begin{center}
    \includegraphics[width=.8\textwidth]{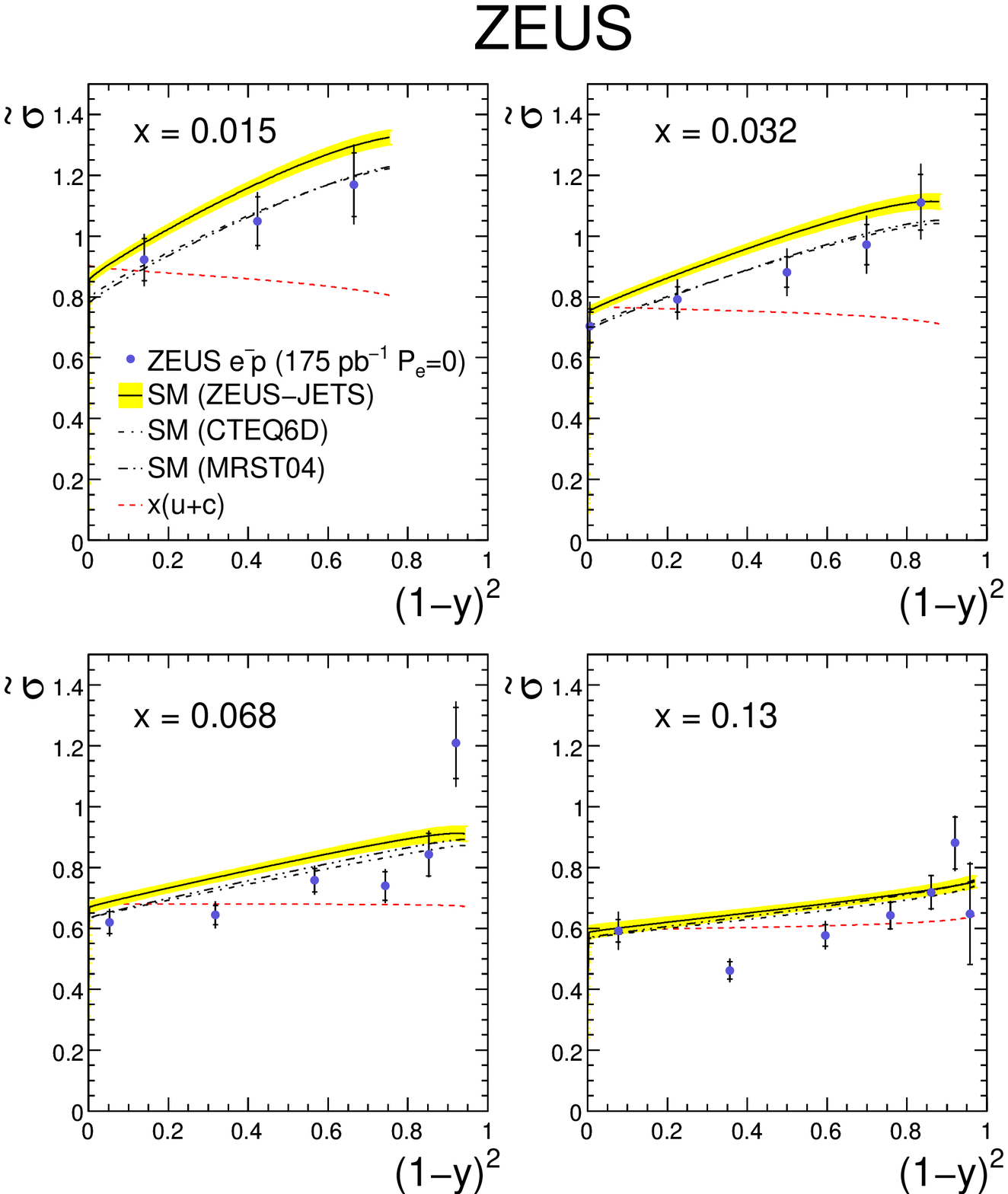}
  \end{center}
  \caption{ 
    The $e^- p$ CC DIS reduced cross section plotted as a function of $(1-y)^2$ 
    for fixed $x$. The circles represent the data points
    and the curves show the predictions of the SM evaluated using
    the ZEUS-JETS, CTEQ6D and MRST04 PDFs.  
    The dashed lines show the contributions of the PDF 
    combination $x(u+c)$ and the shaded band shows the experimental 
    uncertainty from the ZEUS-JETS fit.}
  \label{f:helicity}
\end{figure}

The $W$ boson couples only to left-handed fermions and right-handed anti-fermions. Therefore, the angular 
distribution of the scattered quark in $e^{-}q$ CC DIS will be flat in the electron-quark 
centre-of-mass scattering angle, $\theta^{*}$, 
while it will exhibit a $(1+\cos\theta^{*})^{2}$ distribution in $e^{-}\bar{q}$ scattering. 
Since $(1-y)^2 \propto (1+\cos\theta^{*})^{2}$, the helicity structure of CC interactions can be illustrated 
by plotting the reduced double-differential cross section 
versus $(1-y)^2$ in bins of $x$. This is shown in Fig.~\ref{f:helicity}.
In the region of approximate scaling, i.e. $x\sim 0.1$,
this yields a straight line. At leading order in QCD, the intercept of this line gives 
the ($u+c$) contribution, while the slope gives the ($\bar{d}+\bar{s}$) 
contribution.

\section{Summary}
\label{s:summary}
The cross sections for charged current 
deep inelastic scattering in $e^{-}p$ collisions with 
longitudinally polarised electron beams have been measured. 
The measurements are based on a data sample with an integrated luminosity 
of 175\pb$^{-1}$ collected with the ZEUS detector at HERA 
at a centre-of-mass energy of 318\gev. 
The total cross section is given for positive and negative 
values of the longitudinal polarisation of the electron beam.
In addition, the differential cross-sections $d\sigma/d Q^2$, $d\sigma/d x$ and $d\sigma/d y$ for 
$Q^{2}>200\gev^2$ and 
$d^2 \sigma/d x d Q^2$ are presented in the kinematic range 
$280<Q^2<30\,000 \gev^2$ and $0.015<x<0.65$.
Overall the measured cross sections are well described by the predictions of 
the Standard Model.

%
\section*{Acknowledgements}
 
We appreciate the contributions to the construction and maintenance
of the ZEUS detector of many people who are not listed as authors.
The HERA machine group and the DESY computing staff are especially
acknowledged for their success in providing excellent operation of the
collider and the data-analysis environment.
We thank the DESY directorate for their strong support and encouragement.
It is also a pleasure to thank H. Spiesberger and W. Hollik for helpful discussions.

\vfill\eject

{
\def\bibname{\Large\bf References}
\def\refname{\Large\bf References}
\pagestyle{plain}
\ifzeusbst
  \bibliographystyle{./BiBTeX/bst/l4z_default}
\fi
\ifzdrftbst
  \bibliographystyle{./BiBTeX/bst/l4z_draft}
\fi
\ifzbstepj
  \bibliographystyle{./BiBTeX/bst/l4z_epj}
\fi
\ifzbstnp
  \bibliographystyle{./BiBTeX/bst/l4z_np}
\fi
\ifzbstpl
  \bibliographystyle{./BiBTeX/bst/l4z_pl}
\fi
{\raggedright
\bibliography{./BiBTeX/user/syn.bib,%
              ./BiBTeX/bib/l4z_articles.bib,%
              ./BiBTeX/bib/l4z_books.bib,%
              ./BiBTeX/bib/l4z_conferences.bib,%
              ./BiBTeX/bib/l4z_h1.bib,%
              ./BiBTeX/bib/l4z_misc.bib,%
              ./BiBTeX/bib/l4z_old.bib,%
              ./BiBTeX/bib/l4z_preprints.bib,%
              ./BiBTeX/bib/l4z_replaced.bib,%
              ./BiBTeX/bib/l4z_temporary.bib,%
              ./BiBTeX/bib/l4z_zeus.bib}}
}
\vfill\eject

\begin{table}[p]

\begin{center}

\begin{tabular}{|c|c|c|}

\hline
$P_{e}$ & $\sigma^{\rm CC}$ (pb)\\
\hline
$-0.32 \pm 0.01$ & $89.7 \pm 2.9 \pm 3.6$ \\ 
$-0.29 \pm 0.01$ & $81.9 \pm 2.9 \pm 3.3$ \\ 
$-0.27 \pm 0.01$ & $82.0 \pm 2.9 \pm 3.3$ \\ 
$-0.23 \pm 0.01$ & $78.2 \pm 2.5 \pm 3.1$ \\ 
$-0.16 \pm 0.01$ & $72.5 \pm 2.6 \pm 2.9$ \\ 
$+0.15 \pm 0.01$ & $58.7 \pm 2.6 \pm 2.7$ \\ 
$+0.25 \pm 0.01$ & $48.4 \pm 2.3 \pm 2.3$ \\ 
$+0.32 \pm 0.01$ & $47.1 \pm 2.3 \pm 2.2$ \\ 
$+0.36 \pm 0.01$ & $38.7 \pm 2.3 \pm 1.8$ \\ 
\hline

\end{tabular}

\caption{Values of the total cross section with statistical and systematic uncertainties.}

\label{t:manybins}

\end{center}

\end{table}

\begin{table}[p]

\begin{center}

\begin{tabular}{|c|c|c|c|}

\hline
$Q^2$ range ($\gev^2$) & $Q^2$ ($\gev^2$) & \multicolumn{2}{c|}{$d\sigma/dQ^{2}$ (pb/$\gev^2$)}\\
\hline
& & $P_{e}=+0.30$ & $P_{e}=-0.27$ \\
\hline
$200 - 400$ & $280$ & $(3.00 \pm 0.25 \pm 0.21)\cdot 10^{-2}$ & $(5.13\pm 0.27$ $^{+0.37} _{-0.36}) \cdot 10^{-2}$ \\
$400 - 711$ & $530$ & $(1.93 \pm 0.13 \pm 0.08)\cdot 10^{-2}$ & $(3.57\pm 0.15\pm 0.15)\cdot 10^{-2}$ \\
$711 - 1265$ & $950$ & $(1.40 \pm 0.08 \pm 0.05)\cdot 10^{-2}$ & $(2.30 \pm 0.08 \pm 0.09)\cdot 10^{-2}$ \\
$1265 - 2249$ & $1700$ & $(8.86 \pm 0.44$ $^{+0.31}_{-0.29})\cdot10^{-3}$ & $(1.49 \pm 0.05$ $^{+0.05} _{-0.04}) \cdot 10^{-2}$ \\
$2249 - 4000$ & $3000$ & $(4.19 \pm 0.22 \pm 0.27)\cdot 10^{-3}$ & $(8.14 \pm 0.26 \pm 0.24) \cdot 10^{-3}$ \\
$4000 - 7113$ & $5300$ & $(1.99 \pm 0.12$ $^{+0.10}_{-0.09})\cdot10^{-3}$ & $(3.46 \pm 0.13 \pm 0.17)\cdot 10^{-3}$ \\
$7113 - 12469$ & $9500$ & $(6.74 \pm 0.53$ $^{+0.61}_{-0.58})\cdot 10^{-4}$ & $(1.34 \pm 0.06 \pm 0.12)\cdot 10^{-3}$ \\
$12469 - 22494$ & $17000$ & $(1.65 \pm 0.19$ $^{+0.25}_{-0.21})\cdot10^{-4}$ & $(2.88 \pm 0.21$ $^{+0.43}_{-0.36})\cdot 10^{-4}$ \\
$22494 - 60000$ & $30000$ & $(3.47 \pm 0.67$ $^{+1.36}_{-1.20})\cdot10^{-5}$ & $(5.50 \pm 0.71$ $^{+1.85}_{-1.54})\cdot 10^{-5}$ \\
\hline
$x$ range & $x$ & \multicolumn{2}{c|}{$d\sigma/dx$ (pb)}\\
\hline
& & $P_{e}=+0.30$ & $P_{e}=-0.27$ \\
\hline
$0.01 - 0.021$ & $0.015$ & $424.6 \pm 33.4$ $^{+24.3} _{-23.4}$ & $730.1 \pm 36.3$ $^{+50.6} _{-49.3}$ \\
$0.021 - 0.046$ & $0.032$ & $302.2 \pm 16.2$ $^{+21.7} _{-21.4}$ & $573.9 \pm 18.6$ $^{+36.1} _{-35.6}$ \\
$0.046 - 0.1$ & $0.068$ & $210.5 \pm 9.2$ $^{+7.7} _{-7.6}$ & $352.1 \pm 9.9$ $^{+12.5} _{-12.3}$ \\
$0.1 - 0.178$ & $0.13$ & $119.5 \pm 5.9 \pm 4.4$ & $202.9 \pm 6.4 \pm 7.8$ \\
$0.237 - 0.316$ & $0.24$ & $53.0 \pm 3.0$ $^{+2.8} _{-2.7}$ & $104.4 \pm 3.5$ $^{+5.1} _{-4.9}$ \\
$0.316 - 0.562$ & $0.42$ & $18.8 \pm 1.5$ $^{+1.9} _{-1.7}$ & $32.0 \pm 1.6$ $^{+3.1} _{-2.8}$ \\
$0.562 - 1$ & $0.65$ & $1.69^{+0.78} _{-0.56}$ $^{+0.49} _{-0.42}$ & $5.33 \pm 0.84$ $^{+1.75} _{-1.56}$ \\
\hline
$y$ range & $y$ & \multicolumn{2}{c|}{$d\sigma/dy$ (pb)}\\
\hline
& & $P_{e}=+0.30$ & $P_{e}=-0.27$ \\
\hline
$0.0 - 0.1$ & $0.05$ & $118.6 \pm 6.4 \pm 5.4$ & $210.2 \pm 7.1$ $^{+10.3} _{-10.2}$ \\
$0.1 - 0.2$ & $0.15$ & $80.2 \pm 3.9 \pm 2.3$ & $137.9 \pm 4.3 \pm 2.7$ \\
$0.2 - 0.34$ & $0.27$ & $57.0 \pm 2.8 \pm 2.1$ & $101.5 \pm 3.1 \pm 3.7$ \\
$0.34 - 0.48$ & $0.41$ & $41.9 \pm 2.5 \pm 1.9$ & $73.7 \pm 2.7 \pm 3.8$ \\
$0.48 - 0.62$ & $0.55$ & $36.4 \pm 2.4 \pm 2.2$ & $62.5 \pm 2.6 \pm 4.2$ \\
$0.62 - 0.76$ & $0.69$ & $27.3 \pm 2.2$ $^{+1.4} _{-1.3}$ & $55.6 \pm 2.6$ $^{+2.8} _{-2.7}$ \\
$0.76 - 0.9$ & $0.83$ & $24.5 \pm 2.4$ $^{+2.2} _{-2.1}$ & $44.6 \pm 2.6$ $^{+4.9} _{-4.7}$ \\
\hline

\end{tabular}

\normalsize

\caption{Values of the differential cross-sections $d \sigma /dQ^{2}$,
  $d \sigma /dx$ and $d \sigma /dy$ for $P_{e}= +0.30 \pm 0.01$ and
  $P_{e}= -0.27 \pm 0.01$. The following quantities are given: the
  range of the measurement; the value at which the cross section is
  quoted and the measured cross section, with statistical and systematic uncertainties.}
  \label{t:single-diff}

\end{center}

\end{table}

\begin{table}[p]

\begin{center}

\begin{tabular}{|c|c|c|c|c|c|}

\hline

\multicolumn{6}{|c|}{$d\sigma/dQ^{2}$}\\

\hline

$Q^2$ ($\gev^2$)     & $d\sigma/dQ^{2}$ (pb/$\gev^2$)& $\delta_{{\rm stat}}$ (\%)&  $\delta_{{\rm syst}}$ (\%)&   $\delta_{{\rm unc}}$ (\%)& $\delta_{\rm es}$ (\%)\\ 

\hline
280 & $3.00 \cdot 10^{-2}$ & $ \pm 8.5$ & $ ^{+7.1} _{-6.9} $ & $ ^{+5.4} _{-5.4} $ & $ ^{+4.6} _{-4.2} $  \\
530 & $1.93 \cdot 10^{-2}$ & $ \pm 7.0$ & $ ^{+4.2} _{-4.2} $ & $ ^{+2.0} _{-2.0} $ & $ ^{+3.7} _{-3.7} $  \\
950 & $1.40 \cdot 10^{-2}$ & $ \pm 5.6$ & $ ^{+3.4} _{-3.5} $ & $ ^{+2.1} _{-2.1} $ & $ ^{+2.6} _{-2.8} $  \\
1700 & $8.86 \cdot 10^{-3}$ & $ \pm 5.0$ & $ ^{+3.5} _{-3.3} $ & $ ^{+3.0} _{-3.0} $ & $ ^{+1.7} _{-1.3} $  \\
3000 & $4.19 \cdot 10^{-3}$ & $ \pm 5.4$ & $ ^{+6.4} _{-6.4} $ & $ ^{+6.4} _{-6.4} $ & $ ^{-0.1} _{+0.2} $  \\
5300 & $1.99 \cdot 10^{-3}$ & $ \pm 5.9$ & $ ^{+4.8} _{-4.7} $ & $ ^{+4.2} _{-4.2} $ & $ ^{-1.9} _{+2.3} $  \\
9500 & $6.74 \cdot 10^{-4}$ & $ \pm 7.8$ & $ ^{+9.0} _{-8.5} $ & $ ^{+7.0} _{-7.0} $ & $ ^{-4.9} _{+5.7} $  \\
17000 & $1.65 \cdot 10^{-4}$ & $ \pm 11.8$ & $ ^{+15.2} _{-12.6} $ & $ ^{+8.1} _{-8.1} $ & $ ^{-9.7} _{+12.9} $  \\
30000 & $3.47 \cdot 10^{-5}$ & $ \pm 19.4$ & $ ^{+39.3} _{-34.5} $ & $ ^{+27.6} _{-27.6} $ & $ ^{-20.6} _{+27.9} $  \\
\hline

\multicolumn{6}{|c|}{$d\sigma/dx$}\\

\hline

$x$      &  $d\sigma/dx$ (pb)& $\delta_{{\rm stat}}$ (\%)&  $\delta_{{\rm syst}}$ (\%)&   $\delta_{{\rm unc}}$ (\%)& $\delta_{\rm es}$ (\%)\\ 

\hline
0.015 & 424.6& $ \pm 7.9$ & $ ^{+5.7} _{-5.5} $ & $ ^{+3.9} _{-3.9} $ & $ ^{+4.2} _{-3.9} $  \\
0.032 & 302.2& $ \pm 5.4$ & $ ^{+7.2} _{-7.1} $ & $ ^{+6.6} _{-6.6} $ & $ ^{+2.7} _{-2.5} $  \\
0.068 & 210.5& $ \pm 4.4$ & $ ^{+3.6} _{-3.6} $ & $ ^{+3.5} _{-3.5} $ & $ ^{+1.2} _{-1.0} $  \\
0.13 & 119.5& $ \pm 4.9$ & $ ^{+3.7} _{-3.7} $ & $ ^{+3.6} _{-3.6} $ & $ ^{-0.2} _{+0.4} $  \\
0.24 & 53.0& $ \pm 5.6$ & $ ^{+5.2} _{-5.1} $ & $ ^{+4.2} _{-4.2} $ & $ ^{-2.8} _{+3.1} $  \\
0.42 & 18.8& $ \pm 7.9$ & $ ^{+9.9} _{-8.9} $ & $ ^{+5.2} _{-5.2} $ & $ ^{-7.3} _{+8.5} $  \\
0.65 & 1.69& $^{+46.1} _{-33.0} $ & $ ^{+29.2} _{-25.0} $ & $ ^{+13.9} _{-13.9} $ & $ ^{-20.8} _{+25.7} $  \\
\hline

\multicolumn{6}{|c|}{$d\sigma/dy$}\\

\hline
 
$y$      &  $d\sigma/dy$ (pb) & $\delta_{{\rm stat}}$ (\%)&  $\delta_{{\rm syst}}$ (\%)&   $\delta_{{\rm unc}}$ (\%)& $\delta_{\rm es}$ (\%)\\ 

\hline
0.05 & 118.6& $ \pm 5.4$ & $ ^{+4.6} _{-4.5} $ & $ ^{+4.3} _{-4.3} $ & $ ^{+1.5} _{-1.4} $  \\
0.15 & 80.2& $ \pm 4.9$ & $ ^{+2.9} _{-2.9} $ & $ ^{+2.8} _{-2.8} $ & $ ^{+0.6} _{-0.7} $  \\
0.27 & 57.0& $ \pm 4.9$ & $ ^{+3.7} _{-3.7} $ & $ ^{+3.6} _{-3.6} $ & $ ^{+0.4} _{-0.4} $  \\
0.41 & 41.9& $ \pm 5.9$ & $ ^{+4.6} _{-4.6} $ & $ ^{+4.6} _{-4.6} $ & $ ^{+0.3} _{+0.0} $  \\
0.55 & 36.4& $ \pm 6.5$ & $ ^{+6.0} _{-6.1} $ & $ ^{+6.0} _{-6.0} $ & $ ^{-0.8} _{+0.4} $  \\
0.69 & 27.3& $ \pm 7.9$ & $ ^{+5.0} _{-4.7} $ & $ ^{+4.5} _{-4.5} $ & $ ^{-1.3} _{+2.0} $  \\
0.83 & 24.5& $ \pm 9.6$ & $ ^{+9.0} _{-8.4} $ & $ ^{+6.5} _{-6.5} $ & $ ^{-5.3} _{+6.2} $  \\
\hline

\end{tabular}

\normalsize

\caption{Values of the differential cross-sections $d \sigma /dQ^{2}$,
  $d \sigma /dx$ and $d \sigma /dy$ for $P_{e}=+0.30 \pm 0.01$. The following quantities are given: the value at which the cross section is quoted; the measured cross section; the statistical uncertainty; the total systematic uncertainty; the uncorrelated systematic uncertainty and the calorimeter energy-scale uncertainty ($\delta_{es}$), which has significant correlations between cross-section bins.}  

\label{t:uncorr_single_pos}

\end{center}

\end{table}

\begin{table}[p]

\begin{center}

\begin{tabular}{|c|c|c|c|c|c|}

\hline

\multicolumn{6}{|c|}{$d\sigma/dQ^{2}$}\\

\hline

$Q^2$ ($\gev^2$)     & $d\sigma/dQ^{2}$ (pb/$\gev^2$)& $\delta_{{\rm stat}}$ (\%)&  $\delta_{{\rm syst}}$ (\%)&   $\delta_{{\rm unc}}$ (\%)& $\delta_{\rm es}$ (\%)\\ 

\hline
280 & $5.13 \cdot 10^{-2}$ & $ \pm 5.4$ & $ ^{+7.2} _{-6.9} $ & $ ^{+5.5} _{-5.5} $ & $ ^{+4.6} _{-4.2} $  \\
530 & $3.57 \cdot 10^{-2}$ & $ \pm 4.3$ & $ ^{+4.1} _{-4.2} $ & $ ^{+1.9} _{-1.9} $ & $ ^{+3.7} _{-3.7} $  \\
950 & $2.30 \cdot 10^{-2}$ & $ \pm 3.6$ & $ ^{+4.0} _{-4.0} $ & $ ^{+2.9} _{-2.9} $ & $ ^{+2.6} _{-2.8} $  \\
1700 & $1.49 \cdot 10^{-2}$ & $ \pm 3.2$ & $ ^{+3.2} _{-3.0} $ & $ ^{+2.6} _{-2.6} $ & $ ^{+1.7} _{-1.3} $  \\
3000 & $8.14 \cdot 10^{-3}$ & $ \pm 3.2$ & $ ^{+3.0} _{-3.0} $ & $ ^{+3.0} _{-3.0} $ & $ ^{-0.1} _{+0.2} $  \\
5300 & $3.46 \cdot 10^{-3}$ & $ \pm 3.7$ & $ ^{+5.0} _{-4.8} $ & $ ^{+4.4} _{-4.4} $ & $ ^{-1.9} _{+2.3} $  \\
9500 & $1.34 \cdot 10^{-3}$ & $ \pm 4.6$ & $ ^{+9.2} _{-8.7} $ & $ ^{+7.2} _{-7.2} $ & $ ^{-4.9} _{+5.7} $  \\
17000 & $2.88 \cdot 10^{-4}$ & $ \pm 7.4$ & $ ^{+15.1} _{-12.5} $ & $ ^{+7.8} _{-7.8} $ & $ ^{-9.7} _{+12.9} $  \\
30000 & $5.50 \cdot 10^{-5}$ & $ \pm 12.8$ & $ ^{+33.6} _{-27.9} $ & $ ^{+18.8} _{-18.8} $ & $ ^{-20.6} _{+27.9} $  \\
\hline

\multicolumn{6}{|c|}{$d\sigma/dx$}\\

\hline

$x$      &  $d\sigma/dx$ (pb)& $\delta_{{\rm stat}}$ (\%)&  $\delta_{{\rm syst}}$ (\%)&   $\delta_{{\rm unc}}$ (\%)& $\delta_{\rm es}$ (\%)\\ 

\hline
0.015 & 730.1& $ \pm 5.0$ & $ ^{+6.9} _{-6.7} $ & $ ^{+5.5} _{-5.5} $ & $ ^{+4.2} _{-3.9} $  \\
0.032 & 573.9& $ \pm 3.2$ & $ ^{+6.3} _{-6.2} $ & $ ^{+5.7} _{-5.7} $ & $ ^{+2.7} _{-2.5} $  \\
0.068 & 352.1& $ \pm 2.8$ & $ ^{+3.6} _{-3.5} $ & $ ^{+3.4} _{-3.4} $ & $ ^{+1.2} _{-1.0} $  \\
0.13 & 202.9& $ \pm 3.1$ & $ ^{+3.9} _{-3.8} $ & $ ^{+3.8} _{-3.8} $ & $ ^{-0.2} _{+0.4} $  \\
0.24 & 104.4& $ \pm 3.3$ & $ ^{+4.9} _{-4.7} $ & $ ^{+3.8} _{-3.8} $ & $ ^{-2.8} _{+3.1} $  \\
0.42 & 32.0& $ \pm 5.0$ & $ ^{+9.8} _{-8.8} $ & $ ^{+4.9} _{-4.9} $ & $ ^{-7.3} _{+8.5} $  \\
0.65 & 5.33& $ \pm 15.8$ & $ ^{+32.9} _{-29.2} $ & $ ^{+20.6} _{-20.6} $ & $ ^{-20.8} _{+25.7} $  \\
\hline

\multicolumn{6}{|c|}{$d\sigma/dy$}\\

\hline

$y$      &  $d\sigma/dy$ (pb) & $\delta_{{\rm stat}}$ (\%)&  $\delta_{{\rm syst}}$ (\%)&   $\delta_{{\rm unc}}$ (\%)& $\delta_{\rm es}$ (\%)\\ 

\hline
0.05 & 210.2& $ \pm 3.4$ & $ ^{+4.9} _{-4.9} $ & $ ^{+4.6} _{-4.6} $ & $ ^{+1.5} _{-1.4} $  \\
0.15 & 137.9& $ \pm 3.1$ & $ ^{+1.9} _{-1.9} $ & $ ^{+1.8} _{-1.8} $ & $ ^{+0.6} _{-0.7} $  \\
0.27 & 101.5& $ \pm 3.1$ & $ ^{+3.6} _{-3.6} $ & $ ^{+3.6} _{-3.6} $ & $ ^{+0.4} _{-0.4} $  \\
0.41 & 73.7& $ \pm 3.7$ & $ ^{+5.2} _{-5.2} $ & $ ^{+5.2} _{-5.2} $ & $ ^{+0.3} _{+0.0} $  \\
0.55 & 62.5& $ \pm 4.1$ & $ ^{+6.7} _{-6.8} $ & $ ^{+6.7} _{-6.7} $ & $ ^{-0.8} _{+0.4} $  \\
0.69 & 55.6& $ \pm 4.6$ & $ ^{+5.1} _{-4.9} $ & $ ^{+4.7} _{-4.7} $ & $ ^{-1.3} _{+2.0} $  \\
0.83 & 44.6& $ \pm 5.9$ & $ ^{+11.0} _{-10.6} $ & $ ^{+9.1} _{-9.1} $ & $ ^{-5.3} _{+6.2} $  \\
\hline

\end{tabular}

\normalsize

\caption{Values of the differential cross-sections $d \sigma /dQ^{2}$,
  $d \sigma /dx$ and $d \sigma /dy$ for $P_{e}=-0.27 \pm 0.01$. The following quantities are given: the value at which the cross section is quoted; the measured cross section; the statistical uncertainty; the total systematic uncertainty; the uncorrelated systematic uncertainty and the calorimeter energy-scale uncertainty ($\delta_{es}$), which has significant correlations between cross-section bins.}  

\label{t:uncorr_single_neg}

\end{center}

\end{table}

\begin{table}[p]

\footnotesize

\begin{center}

\begin{tabular}{|c|c|c|c|c|}

\hline

$Q^2$ ($\gev^2$) & $x$ & \multicolumn{3}{c|}{$\tilde{\sigma}$} \\

\hline

& & $P_{e}=-0.27$ & $P_{e}=+0.30$ &  $P_{e}=0$\\

\hline
280 & 0.015 & $1.39 \pm 0.14 \pm 0.11$ & $1.03 \pm 0.15 \pm 0.07$ & $1.17 \pm 0.10 \pm 0.08$ \\
280 & 0.032 & $1.54 \pm 0.14$ $^{+0.12} _{-0.11}$ & $0.65 \pm 0.11 \pm 0.05$ & $1.11 \pm 0.09$ $^{+0.09} _{-0.08}$ \\
280 & 0.068 & $1.54 \pm 0.17$ $^{+0.10} _{-0.11}$ & $0.91 \pm 0.16$ $^{+0.05} _{-0.06}$ & $1.21 \pm 0.12$ $^{+0.07} _{-0.08}$ \\
280 & 0.13 & $0.68^{+0.29} _{-0.21}$ $^{+0.05} _{-0.03}$ & $0.69^{+0.37} _{-0.26}$ $^{+0.05} _{-0.03}$ & $0.65 \pm 0.17$ $^{+0.04} _{-0.03}$ \\
530 & 0.015 & $1.46 \pm 0.12$ $^{+0.10} _{-0.09}$ & $0.61 \pm 0.09 \pm 0.03$ & $1.05 \pm 0.08 \pm 0.05$ \\
530 & 0.032 & $1.28 \pm 0.10 \pm 0.08$ & $0.67 \pm 0.09 \pm 0.05$ & $0.97 \pm 0.07 \pm 0.07$ \\
530 & 0.068 & $1.04 \pm 0.10 \pm 0.04$ & $0.69 \pm 0.10$ $^{+0.03} _{-0.02}$ & $0.84 \pm 0.07$ $^{+0.04} _{-0.03}$ \\
530 & 0.13 & $1.11 \pm 0.12$ $^{+0.04} _{-0.05}$ & $0.69 \pm 0.12$ $^{+0.02} _{-0.03}$ & $0.88 \pm 0.09$ $^{+0.03} _{-0.04}$ \\
950 & 0.015 & $1.14 \pm 0.10$ $^{+0.08} _{-0.09}$ & $0.74 \pm 0.10 \pm 0.04$ & $0.92 \pm 0.07 \pm 0.05$ \\
950 & 0.032 & $1.15 \pm 0.07 \pm 0.08$ & $0.62 \pm 0.06 \pm 0.04$ & $0.88 \pm 0.05 \pm 0.06$ \\
950 & 0.068 & $0.92 \pm 0.07 \pm 0.04$ & $0.59 \pm 0.07 \pm 0.02$ & $0.74 \pm 0.05 \pm 0.03$ \\
950 & 0.13 & $0.88 \pm 0.08 \pm 0.03$ & $0.59 \pm 0.08$ $^{+0.01} _{-0.02}$ & $0.72 \pm 0.05 \pm 0.02$ \\
950 & 0.24 & $0.68 \pm 0.08 \pm 0.02$ & $0.40 \pm 0.08 \pm 0.01$ & $0.53 \pm 0.06 \pm 0.02$ \\
1700 & 0.032 & $1.00 \pm 0.06 \pm 0.06$ & $0.60 \pm 0.06 \pm 0.04$ & $0.79 \pm 0.04 \pm 0.05$ \\
1700 & 0.068 & $0.94 \pm 0.05 \pm 0.04$ & $0.60 \pm 0.05 \pm 0.03$ & $0.76 \pm 0.04 \pm 0.03$ \\
1700 & 0.13 & $0.83 \pm 0.06 \pm 0.03$ & $0.47 \pm 0.06$ $^{+0.02} _{-0.01}$ & $0.64 \pm 0.04 \pm 0.02$ \\
1700 & 0.24 & $0.68 \pm 0.06 \pm 0.02$ & $0.38 \pm 0.05 \pm 0.01$ & $0.53 \pm 0.04 \pm 0.01$ \\
1700 & 0.42 & $0.41 \pm 0.07 \pm 0.02$ & $0.19 \pm 0.05 \pm 0.01$ & $0.30 \pm 0.05$ $^{+0.02} _{-0.01}$ \\
3000 & 0.032 & $0.96 \pm 0.08 \pm 0.07$ & $0.44 \pm 0.07 \pm 0.04$ & $0.70 \pm 0.06 \pm 0.06$ \\
3000 & 0.068 & $0.84 \pm 0.05 \pm 0.03$ & $0.46 \pm 0.04 \pm 0.03$ & $0.64 \pm 0.03 \pm 0.03$ \\
3000 & 0.13 & $0.72 \pm 0.05 \pm 0.03$ & $0.45 \pm 0.05 \pm 0.03$ & $0.58 \pm 0.04 \pm 0.03$ \\
3000 & 0.24 & $0.68 \pm 0.05 \pm 0.02$ & $0.31 \pm 0.04 \pm 0.02$ & $0.50 \pm 0.03 \pm 0.02$ \\
3000 & 0.42 & $0.33 \pm 0.04 \pm 0.02$ & $0.12 \pm 0.03 \pm 0.01$ & $0.23 \pm 0.02$ $^{+0.02} _{-0.01}$ \\
5300 & 0.068 & $0.80 \pm 0.05 \pm 0.04$ & $0.45 \pm 0.05 \pm 0.02$ & $0.62 \pm 0.04 \pm 0.03$ \\
5300 & 0.13 & $0.59 \pm 0.04 \pm 0.04$ & $0.34 \pm 0.04 \pm 0.02$ & $0.46 \pm 0.03 \pm 0.03$ \\
5300 & 0.24 & $0.55 \pm 0.04 \pm 0.03$ & $0.29 \pm 0.04 \pm 0.02$ & $0.42 \pm 0.03 \pm 0.02$ \\
5300 & 0.42 & $0.30 \pm 0.03 \pm 0.02$ & $0.22 \pm 0.03 \pm 0.02$ & $0.25 \pm 0.02 \pm 0.02$ \\
5300 & 0.65 & $0.11 \pm 0.03 \pm 0.03$ & & $0.07 \pm 0.02$ $^{+0.01} _{-0.02}$ \\
9500 & 0.13 & $0.81 \pm 0.06 \pm 0.07$ & $0.37 \pm 0.05 \pm 0.03$ & $0.59 \pm 0.04 \pm 0.05$ \\
9500 & 0.24 & $0.53 \pm 0.04$ $^{+0.06} _{-0.05}$ & $0.27 \pm 0.04 \pm 0.03$ & $0.40 \pm 0.03 \pm 0.04$ \\
9500 & 0.42 & $0.26 \pm 0.03$ $^{+0.03} _{-0.02}$ & $0.14 \pm 0.03 \pm 0.01$ & $0.20 \pm 0.02 \pm 0.02$ \\
9500 & 0.65 & $0.03^{+0.02} _{-0.01} \pm 0.01$ & $0.03^{+0.03} _{-0.02} \pm 0.01$ & $0.03^{+0.02} _{-0.01} \pm 0.01$ \\
17000 & 0.24 & $0.52 \pm 0.05 \pm 0.07$ & $0.25 \pm 0.04$ $^{+0.04} _{-0.03}$ & $0.39 \pm 0.03 \pm 0.05$ \\
17000 & 0.42 & $0.20 \pm 0.03 \pm 0.03$ & $0.14 \pm 0.03 \pm 0.02$ & $0.17 \pm 0.02$ $^{+0.03} _{-0.02}$ \\
17000 & 0.65 & $0.05^{+0.03} _{-0.02}$ $^{+0.02} _{-0.01}$ & & $0.03^{+0.02} _{-0.01} \pm 0.01$ \\
30000 & 0.42 & $0.26 \pm 0.04$ $^{+0.09} _{-0.07}$ & $0.21 \pm 0.04$ $^{+0.08} _{-0.07}$ & $0.23 \pm 0.03$ $^{+0.08} _{-0.07}$ \\
30000 & 0.65 & $0.05 \pm 0.02$ $^{+0.03} _{-0.02}$ & & $0.03^{+0.02} _{-0.01}$ $^{+0.02} _{-0.01}$ \\
\hline

\end{tabular}

\caption{Values of the reduced cross sections. The following quantities
  are given: the values of $Q^2$ and $x$ at which the cross section is
  quoted and the measured cross section, with statistical and systematic uncertainties.
  Three bins in the $P_e=+0.30$ cross section were judged to be too statistically imprecise 
  to be quoted without combination with the $P_e=-0.27$ data and are therefore omitted from the table.}
\label{t:double}

\end{center}

\end{table}

\begin{table}[p]

\footnotesize

\begin{center}

\begin{tabular}{|c|c|c|c|c|c|c|}

\hline

$Q^2$ ($\gev^2$)     & $x$ &$\tilde{\sigma}$ & $\delta_{{\rm stat}}$ (\%)&  $\delta_{{\rm syst}}$ (\%)&   $\delta_{{\rm unc}}$ (\%)& $\delta_{\rm es}$ (\%)\\ 

\hline
280 & 0.015 & 1.03& $ \pm 14.7$ & $ ^{+7.2} _{-7.1} $ & $ ^{+5.7} _{-5.7} $ & $ ^{+4.3} _{-4.1} $  \\
280 & 0.032 & 0.65& $ \pm 16.7$ & $ ^{+8.3} _{-7.7} $ & $ ^{+7.3} _{-7.3} $ & $ ^{+4.0} _{-2.4} $  \\
280 & 0.068 & 0.91& $ \pm 17.2$ & $ ^{+5.9} _{-6.9} $ & $ ^{+3.7} _{-3.7} $ & $ ^{+4.6} _{-5.8} $  \\
280 & 0.13 & 0.69& $^{+54.0} _{-37.1} $ & $ ^{+6.5} _{-3.9} $ & $ ^{+3.5} _{-3.5} $ & $ ^{+5.5} _{-1.9} $  \\
530 & 0.015 & 0.61& $ \pm 15.4$ & $ ^{+5.2} _{-4.8} $ & $ ^{+2.9} _{-2.9} $ & $ ^{+4.3} _{-3.8} $  \\
530 & 0.032 & 0.67& $ \pm 12.6$ & $ ^{+7.4} _{-7.5} $ & $ ^{+6.5} _{-6.5} $ & $ ^{+3.4} _{-3.7} $  \\
530 & 0.068 & 0.69& $ \pm 14.1$ & $ ^{+4.4} _{-3.6} $ & $ ^{+2.1} _{-2.1} $ & $ ^{+3.9} _{-2.9} $  \\
530 & 0.13 & 0.69& $ \pm 16.8$ & $ ^{+3.1} _{-4.2} $ & $ ^{+1.1} _{-1.1} $ & $ ^{+2.9} _{-4.0} $  \\
950 & 0.015 & 0.74& $ \pm 12.9$ & $ ^{+5.2} _{-5.9} $ & $ ^{+4.2} _{-4.2} $ & $ ^{+3.1} _{-4.2} $  \\
950 & 0.032 & 0.62& $ \pm 10.4$ & $ ^{+7.2} _{-7.3} $ & $ ^{+6.6} _{-6.6} $ & $ ^{+2.9} _{-3.1} $  \\
950 & 0.068 & 0.59& $ \pm 11.0$ & $ ^{+4.1} _{-4.0} $ & $ ^{+3.0} _{-3.0} $ & $ ^{+2.8} _{-2.7} $  \\
950 & 0.13 & 0.59& $ \pm 12.9$ & $ ^{+2.0} _{-3.0} $ & $ ^{+1.1} _{-1.1} $ & $ ^{+1.7} _{-2.8} $  \\
950 & 0.24 & 0.40& $ \pm 19.4$ & $ ^{+3.5} _{-2.9} $ & $ ^{+2.3} _{-2.9} $ & $ ^{+1.7} _{+1.0} $  \\
1700 & 0.032 & 0.60& $ \pm 9.3$ & $ ^{+7.1} _{-7.1} $ & $ ^{+6.9} _{-6.9} $ & $ ^{+1.8} _{-1.7} $  \\
1700 & 0.068 & 0.60& $ \pm 8.7$ & $ ^{+4.4} _{-4.2} $ & $ ^{+4.1} _{-4.1} $ & $ ^{+1.7} _{-1.1} $  \\
1700 & 0.13 & 0.47& $ \pm 12.2$ & $ ^{+3.8} _{-3.2} $ & $ ^{+2.6} _{-2.6} $ & $ ^{+2.8} _{-1.8} $  \\
1700 & 0.24 & 0.38& $ \pm 12.9$ & $ ^{+3.4} _{-3.5} $ & $ ^{+3.4} _{-3.3} $ & $ ^{-0.3} _{-0.9} $  \\
1700 & 0.42 & 0.19& $ \pm 29.0$ & $ ^{+5.7} _{-5.3} $ & $ ^{+5.0} _{-5.0} $ & $ ^{-2.0} _{+2.8} $  \\
3000 & 0.032 & 0.44& $ \pm 15.6$ & $ ^{+9.7} _{-9.6} $ & $ ^{+9.4} _{-9.4} $ & $ ^{+2.5} _{-1.7} $  \\
3000 & 0.068 & 0.46& $ \pm 8.8$ & $ ^{+7.0} _{-7.0} $ & $ ^{+7.0} _{-7.0} $ & $ ^{+0.0} _{-0.6} $  \\
3000 & 0.13 & 0.45& $ \pm 10.6$ & $ ^{+7.2} _{-7.2} $ & $ ^{+7.2} _{-7.2} $ & $ ^{+0.2} _{+0.6} $  \\
3000 & 0.24 & 0.31& $ \pm 12.2$ & $ ^{+6.4} _{-6.4} $ & $ ^{+6.3} _{-6.3} $ & $ ^{-0.9} _{+0.5} $  \\
3000 & 0.42 & 0.12& $ \pm 21.4$ & $ ^{+8.9} _{-7.8} $ & $ ^{+7.3} _{-7.3} $ & $ ^{-2.8} _{+5.2} $  \\
5300 & 0.068 & 0.45& $ \pm 10.6$ & $ ^{+5.0} _{-4.7} $ & $ ^{+4.6} _{-4.6} $ & $ ^{-1.0} _{+2.0} $  \\
5300 & 0.13 & 0.34& $ \pm 11.0$ & $ ^{+6.1} _{-6.0} $ & $ ^{+6.0} _{-6.0} $ & $ ^{-0.8} _{+1.3} $  \\
5300 & 0.24 & 0.29& $ \pm 12.6$ & $ ^{+5.3} _{-5.3} $ & $ ^{+4.7} _{-4.7} $ & $ ^{-2.6} _{+2.6} $  \\
5300 & 0.42 & 0.22& $ \pm 13.9$ & $ ^{+7.2} _{-7.6} $ & $ ^{+5.0} _{-5.0} $ & $ ^{-5.7} _{+5.2} $  \\
9500 & 0.13 & 0.37& $ \pm 12.2$ & $ ^{+8.3} _{-8.6} $ & $ ^{+7.2} _{-7.2} $ & $ ^{-4.6} _{+4.1} $  \\
9500 & 0.24 & 0.27& $ \pm 13.4$ & $ ^{+11.1} _{-10.2} $ & $ ^{+9.5} _{-9.5} $ & $ ^{-3.9} _{+5.8} $  \\
9500 & 0.42 & 0.14& $ \pm 18.3$ & $ ^{+10.4} _{-9.6} $ & $ ^{+6.6} _{-6.6} $ & $ ^{-7.0} _{+8.1} $  \\
9500 & 0.650 & 0.03& $^{+97.4} _{-54.6} $ & $ ^{+26.4} _{-21.3} $ & $ ^{+14.5} _{-14.5} $ & $ ^{-15.7} _{+22.1} $  \\
17000 & 0.24 & 0.25& $ \pm 15.9$ & $ ^{+14.2} _{-12.9} $ & $ ^{+8.9} _{-8.9} $ & $ ^{-9.3} _{+11.1} $  \\
17000 & 0.42 & 0.14& $ \pm 21.4$ & $ ^{+16.8} _{-14.0} $ & $ ^{+10.0} _{-10.0} $ & $ ^{-9.8} _{+13.5} $  \\
30000 & 0.42 & 0.21& $ \pm 20.9$ & $ ^{+38.3} _{-34.4} $ & $ ^{+28.4} _{-28.4} $ & $ ^{-19.5} _{+25.7} $  \\
\hline

\end{tabular}

\caption{Values of the reduced cross section for $P_{e}=+0.30 \pm 0.01$. The following quantities are given: the values of $Q^2$ and $x$ at which the cross section is quoted; the measured cross section; the statistical uncertainty; the total systematic uncertainty; the uncorrelated systematic uncertainty and the calorimeter energy-scale uncertainty ($\delta_{es}$), which has significant correlations between cross-section bins.}

  \label{t:uncorr_double_pos}

\end{center}

\end{table}

\begin{table}[p]

\footnotesize

\begin{center}

\begin{tabular}{|c|c|c|c|c|c|c|}

\hline

$Q^2$ ($\gev^2$)     & $x$ &$\tilde{\sigma}$ & $\delta_{{\rm stat}}$ (\%)&  $\delta_{{\rm syst}}$ (\%)&   $\delta_{{\rm unc}}$ (\%)& $\delta_{\rm es}$ (\%)\\ 

\hline
280 & 0.015 & 1.39& $ \pm 10.4$ & $ ^{+8.2} _{-8.1} $ & $ ^{+7.0} _{-7.0} $ & $ ^{+4.3} _{-4.1} $  \\
280 & 0.032 & 1.54& $ \pm 8.9$ & $ ^{+7.8} _{-7.1} $ & $ ^{+6.7} _{-6.7} $ & $ ^{+4.0} _{-2.4} $  \\
280 & 0.068 & 1.54& $ \pm 11.0$ & $ ^{+6.3} _{-7.2} $ & $ ^{+4.2} _{-4.2} $ & $ ^{+4.6} _{-5.8} $  \\
280 & 0.13 & 0.68& $^{+42.9} _{-31.3} $ & $ ^{+7.0} _{-4.7} $ & $ ^{+4.3} _{-4.3} $ & $ ^{+5.5} _{-1.9} $  \\
530 & 0.015 & 1.46& $ \pm 8.3$ & $ ^{+6.7} _{-6.4} $ & $ ^{+5.1} _{-5.1} $ & $ ^{+4.3} _{-3.8} $  \\
530 & 0.032 & 1.28& $ \pm 7.6$ & $ ^{+6.5} _{-6.7} $ & $ ^{+5.5} _{-5.5} $ & $ ^{+3.4} _{-3.7} $  \\
530 & 0.068 & 1.04& $ \pm 9.6$ & $ ^{+4.3} _{-3.4} $ & $ ^{+1.8} _{-1.8} $ & $ ^{+3.9} _{-2.9} $  \\
530 & 0.13 & 1.11& $ \pm 11.1$ & $ ^{+3.3} _{-4.3} $ & $ ^{+1.6} _{-1.6} $ & $ ^{+2.9} _{-4.0} $  \\
950 & 0.015 & 1.14& $ \pm 8.7$ & $ ^{+6.9} _{-7.5} $ & $ ^{+6.2} _{-6.2} $ & $ ^{+3.1} _{-4.2} $  \\
950 & 0.032 & 1.15& $ \pm 6.3$ & $ ^{+6.7} _{-6.7} $ & $ ^{+6.0} _{-6.0} $ & $ ^{+2.9} _{-3.1} $  \\
950 & 0.068 & 0.92& $ \pm 7.4$ & $ ^{+4.5} _{-4.5} $ & $ ^{+3.6} _{-3.6} $ & $ ^{+2.8} _{-2.7} $  \\
950 & 0.13 & 0.88& $ \pm 8.8$ & $ ^{+3.1} _{-3.9} $ & $ ^{+2.6} _{-2.6} $ & $ ^{+1.7} _{-2.8} $  \\
950 & 0.24 & 0.68& $ \pm 12.4$ & $ ^{+3.7} _{-3.1} $ & $ ^{+2.5} _{-3.1} $ & $ ^{+1.7} _{+1.0} $  \\
1700 & 0.032 & 1.00& $ \pm 6.0$ & $ ^{+6.1} _{-6.1} $ & $ ^{+5.8} _{-5.8} $ & $ ^{+1.8} _{-1.7} $  \\
1700 & 0.068 & 0.94& $ \pm 5.8$ & $ ^{+4.1} _{-3.9} $ & $ ^{+3.7} _{-3.7} $ & $ ^{+1.7} _{-1.1} $  \\
1700 & 0.13 & 0.82& $ \pm 7.7$ & $ ^{+3.7} _{-3.1} $ & $ ^{+2.5} _{-2.5} $ & $ ^{+2.8} _{-1.8} $  \\
1700 & 0.24 & 0.68& $ \pm 8.0$ & $ ^{+2.5} _{-2.7} $ & $ ^{+2.5} _{-2.4} $ & $ ^{-0.3} _{-0.9} $  \\
1700 & 0.42 & 0.41& $ \pm 16.4$ & $ ^{+5.2} _{-4.8} $ & $ ^{+4.4} _{-4.4} $ & $ ^{-2.0} _{+2.8} $  \\
3000 & 0.032 & 0.96& $ \pm 8.7$ & $ ^{+7.2} _{-6.9} $ & $ ^{+6.7} _{-6.7} $ & $ ^{+2.5} _{-1.7} $  \\
3000 & 0.068 & 0.84& $ \pm 5.4$ & $ ^{+4.0} _{-4.1} $ & $ ^{+4.0} _{-4.0} $ & $ ^{+0.0} _{-0.6} $  \\
3000 & 0.13 & 0.72& $ \pm 7.0$ & $ ^{+4.7} _{-4.6} $ & $ ^{+4.6} _{-4.6} $ & $ ^{+0.2} _{+0.6} $  \\
3000 & 0.24 & 0.68& $ \pm 6.9$ & $ ^{+2.3} _{-2.5} $ & $ ^{+2.3} _{-2.3} $ & $ ^{-0.9} _{+0.5} $  \\
3000 & 0.42 & 0.33& $ \pm 10.7$ & $ ^{+6.7} _{-5.1} $ & $ ^{+4.3} _{-4.3} $ & $ ^{-2.8} _{+5.2} $  \\
5300 & 0.068 & 0.80& $ \pm 6.6$ & $ ^{+5.1} _{-4.8} $ & $ ^{+4.7} _{-4.7} $ & $ ^{-1.0} _{+2.0} $  \\
5300 & 0.13 & 0.59& $ \pm 7.0$ & $ ^{+6.3} _{-6.3} $ & $ ^{+6.2} _{-6.2} $ & $ ^{-0.8} _{+1.3} $  \\
5300 & 0.24 & 0.55& $ \pm 7.6$ & $ ^{+5.1} _{-5.1} $ & $ ^{+4.4} _{-4.4} $ & $ ^{-2.6} _{+2.6} $  \\
5300 & 0.42 & 0.30& $ \pm 10.0$ & $ ^{+7.1} _{-7.5} $ & $ ^{+4.8} _{-4.8} $ & $ ^{-5.7} _{+5.2} $  \\
5300 & 0.65 & 0.11& $ \pm 27.3$ & $ ^{+24.1} _{-25.1} $ & $ ^{+20.5} _{-20.5} $ & $ ^{-14.5} _{+12.7} $  \\
9500 & 0.13 & 0.81& $ \pm 6.9$ & $ ^{+8.6} _{-8.8} $ & $ ^{+7.5} _{-7.5} $ & $ ^{-4.6} _{+4.1} $  \\
9500 & 0.24 & 0.53& $ \pm 8.1$ & $ ^{+11.1} _{-10.2} $ & $ ^{+9.4} _{-9.4} $ & $ ^{-3.9} _{+5.8} $  \\
9500 & 0.42 & 0.26& $ \pm 11.3$ & $ ^{+10.4} _{-9.6} $ & $ ^{+6.6} _{-6.6} $ & $ ^{-7.0} _{+8.1} $  \\
9500 & 0.65 & 0.03& $^{+67.8} _{-43.4} $ & $ ^{+30.5} _{-26.2} $ & $ ^{+21.0} _{-21.0} $ & $ ^{-15.7} _{+22.1} $  \\
17000 & 0.24 & 0.52& $ \pm 9.3$ & $ ^{+14.0} _{-12.6} $ & $ ^{+8.5} _{-8.5} $ & $ ^{-9.3} _{+11.1} $  \\
17000 & 0.42 & 0.20& $ \pm 14.5$ & $ ^{+16.6} _{-13.8} $ & $ ^{+9.6} _{-9.6} $ & $ ^{-9.8} _{+13.5} $  \\
17000 & 0.65 & 0.05& $^{+54.1} _{-37.2} $ & $ ^{+31.6} _{-28.9} $ & $ ^{+20.6} _{-20.6} $ & $ ^{-20.3} _{+23.9} $  \\
30000 & 0.42 & 0.26& $ \pm 15.5$ & $ ^{+32.4} _{-27.8} $ & $ ^{+19.8} _{-19.8} $ & $ ^{-19.5} _{+25.7} $  \\
30000 & 0.65 & 0.05& $^{+50.1} _{-35.2} $ & $ ^{+52.5} _{-40.8} $ & $ ^{+27.7} _{-27.7} $ & $ ^{-30.0} _{+44.6} $  \\
\hline

\end{tabular}

\caption{Values of the reduced cross section for $P_{e}=-0.27 \pm 0.01$. The following quantities are given: the values of $Q^2$ and $x$ at which the cross section is quoted; the measured cross section; the statistical uncertainty; the total systematic uncertainty; the uncorrelated systematic uncertainty and the calorimeter energy-scale uncertainty ($\delta_{es}$), which has significant correlations between cross-section bins.}

  \label{t:uncorr_double_neg}

\end{center}

\end{table}

%
%
\newpage

%
%
\end{document}